\documentclass[12pt]{article}
\usepackage{epsfig}
\usepackage{color}
\usepackage{amssymb,amsmath}
\usepackage{graphicx}
\usepackage{here}
\newcommand{\ba}{\begin{align}}
\newcommand{\ea}{\end{align}}

 \makeatletter
\def\alt{\mathrel{\mathpalette\gl@align<}}
\def\agt{\mathrel{\mathpalette\gl@align>}}
\def\gl@align#1#2{\lower.6ex\vbox{\baselineskip\z@skip\lineskip\z@
\ialign{$\m@th#1\hfil##\hfil$\crcr#2\crcr\sim\crcr}}} \makeatother

\begin{document}
\begin{flushright}
\end{flushright}
\vspace*{1.0cm}

\begin{center}
\baselineskip 20pt 
{\Large\bf 
 Strong dynamics in a classically scale invariant extension \\
 of the Standard Model with flatland
}
\vspace{1cm}

{\large 
Naoyuki Haba \ and \ Toshifumi Yamada
} \vspace{.5cm}

{\baselineskip 20pt \it
Graduate School of Science and Engineering, Shimane University, Matsue 690-8504, Japan
}

\vspace{.5cm}

\vspace{1.5cm} {\bf Abstract} \end{center}

We investigate the scenario where the Standard Model is extended with classical scale invariance, which is broken by chiral symmetry breaking and confinement in a new strongly-coupled gauge theory that resembles QCD.
The Standard Model Higgs field emerges as a result of the mixing of a scalar meson in the new strong dynamics and a massless elementary scalar field.
The mass and scalar decay constant of that scalar meson, which are generated dynamically in the new gauge theory,
 give rise to the Higgs field mass term, automatically possessing the correct negative sign by the bosonic seesaw mechanism.
Using analogy with QCD, we evaluate the dynamical scale of the new gauge theory and further make quantitative predictions for light pseudo-Nambu-Goldstone bosons associated with the spontaneous breaking of axial symmetry along chiral symmetry breaking in the new gauge theory.
A prominent consequence of the scenario is that there should be a Standard Model gauge singlet pseudo-Nambu-Goldstone boson with mass below 220~GeV,
 which couples to two electroweak gauge bosons through the Wess-Zumino-Witten term, whose strength is thus determined by the dynamical scale of the new gauge theory.
Other pseudo-Nambu-Goldstone bosons, charged under the electroweak gauge groups, also appear.
Concerning the theoretical aspects,
 it is shown that the scalar quartic coupling can vanish at the Planck scale with the top quark pole mass as large as 172.5~GeV, 
 realizing the flatland scenario without being in tension with the current experimental data.

\thispagestyle{empty}

\newpage

\setcounter{footnote}{0}
\baselineskip 18pt
%

\section{Introduction}

Now that the Standard Model (SM) picture of electroweak symmetry breaking is soundly supported by the Higgs particle measurement,
 the next theoretical challenge is to derive the Higgs potential, which has been added to the SM \textit{ad hoc} to trigger electroweak symmetry breaking, from an underlying theory behind the SM.
An appealing scenario is that a new strongly-coupled gauge theory generates the Higgs field mass term through dimensional transmutation, 
 as it is capable of addressing the origin of the electroweak scale.
Such scenarios should come along with a symmetry or an assumption that forbids a tree-level Higgs field mass, like classical scale invariance~\cite{csi}, 
 since otherwise the Higgs mass would arise as a genuine parameter of the fundamental Lagrangian, not as a consequence of dynamics.
In this paper, we aim at a derivation of the SM Higgs field mass term from strong dynamics of a new gauge theory in a classically scale invariant setup.
To this end, we consider the minimal model described below:
Besides the SM fermions and gauge fields, our model includes an elementary scalar field with the same electroweak charge as the SM Higgs field, which however is massless due to classical scale invariance.
The model further contains a new $SU(N)_T$ gauge group and 3~flavors of massless Dirac fermions that
 are charged under both the $SU(N)_T$ and electroweak gauge groups and have a Yukawa-type coupling with the elementary scalar field.
At infrared scales, the new gauge theory becomes strongly-coupled and triggers chiral symmetry breaking and confinement, analogously to QCD.
The strong dynamics gives rise to a massive scalar meson as a bound state of the new fermions, which mixes with the elementary scalar through the Yukawa-type coupling.
One of the mass eigenstates necessarily has a negative mass squared term due to the "bosonic seesaw" mechanism~\cite{bosonicseesaw}, at a scale given by the product of the Yukawa-type coupling and the dynamical scale of the new gauge theory.
It is this state that we identify with the SM Higgs field.
We thereby attribute the scale and negative sign of the SM Higgs mass term to dynamics of the new gauge theory.
Since the principal motivation of this study is to link the Higgs field mass to a new dynamical scale,
 we do not adhere to "naturalness" of the electroweak scale;
 we take into account all situations where the dynamical scale is arbitrarily high, as long as the SM Higgs field mass term is reproduced at the correct scale.

Concerning experimental signatures of the model, we intensively study pseudo-Nambu-Goldstone (pNG) bosons associated with
 the spontaneous breaking of $SU(3)$ axial symmetry along chiral symmetry breaking in the $SU(N)_T$ gauge theory.
The pNG bosons acquire mass from the electroweak interactions and the Yukawa-type interaction that explicitly violate the $SU(3)$ axial symmetry,
 which are the only source for their mass because the current mass is zero due to classical scale invariance.
A prominent prediction of the model is that there should be a SM gauge singlet pNG boson with mass below 220~GeV, which couples to the SM $W$, $Z$ bosons and photon through the Wess-Zumino-Witten term~\cite{wz,w}.
The mass prediction is insensitive to model parameters because this pNG boson is massive due only to the Yukawa-type interaction, and hence its mass is given by the dynamical scale times the coupling constant for the Yukawa-type interaction.
Since this is also the scale at which the Higgs field mass term is generated, the pNG boson necessarily has mass at the electroweak scale.
The coupling with the electroweak gauge bosons is, however, suppressed by the unknown dynamical scale of the new gauge theory,
 but searching for a particle with $O(100)$~GeV mass that interacts feebly with electroweak gauge bosons inspires new experimental techniques and is intriguing in its own light.
There also appear pNG bosons with electroweak charges, which gain mass from the electroweak interactions at a scale of the dynamical scale times the electroweak gauge couplings.
Their masses are therefore enhanced accordingly with the dynamical scale and these pNG bosons may not be kinematically accessible at colliders.

The model has an implication for physics at the Planck scale.
The vanishing of the scalar quartic coupling at the Planck scale in a classically scale invariant model, or equivalently, a flat scalar potential at the Planck scale, has been pursued in Ref.~\cite{flatland} in the hope of finding hints for quantum gravity.
In previous models, the vanishing of the Higgs quartic coupling is achieved only with the top quark pole mass well below 172~GeV,
 and hence those models are in tension with top quark mass measurements;
 note that the ATLAS Collaboration has recently reported the top quark mass to be $172.84\pm0.70$~GeV~\cite{topatlas}.
Our model, in contrast, can realize the vanishing of the scalar quartic coupling with the top quark pole mass as large as 172.5~GeV.
This is possible owing to contributions of the new fermions to the renormalization group (RG) running of the weak gauge coupling,
 which enhance this gauge coupling at ultraviolet scales and eventually lift the scalar quartic coupling at the Planck scale through the RG evolution as compared to the SM.

Models where strong dynamics supplies the SM Higgs field mass term in a classically scale invariant setup
 have been proposed in Ref.~\cite{csi-chsb}, which however cannot address the origin of its tachyonic feature.
A model similar to ours has been studied in Refs.~\cite{similar1,similar2,similar3}.
Our model is more attractive in that the field content is minimal and thus a more definite prediction for the pNG boson mass spectrum is feasible.

The content of the paper is as follows:
In Secion~2, we describe the model and expound how the SM Higgs field mass term is generated from strong dynamics of the $SU(N)_T$ gauge theory in the classically scale invariant setup.
We further evaluate the dynamical scale of the gauge theory based on analogy with QCD.
In Section~3, we study phenomenology of the pNG bosons by deriving their mass spectrum and interaction strengths, again relying on analogy with QCD.
In Section~4, we demonstrate that the model can achieve the vanishing of the scalar quartic coupling at the Planck scale with the top quark pole mass as large as 172.5~GeV.
Section~5 is for the summary and discussions.
\\

\section{Model}

\subsection{Model description and the origin of the Standard Model Higgs field mass}

The gauge symmetry of the model is $SU(N)_T\times SU(3)_C\times SU(2)_W\times U(1)_Y$, where
 $SU(3)_C$, $SU(2)_W$ and $U(1)_Y$ are the SM color, weak and hypercharge gauge groups, respectively, and $SU(N)_T$ is a newly introduced gauge group.
We require $N\geq3$.
The model includes the same matter fermions as the SM and further contains massless Dirac fermions, $\chi,\psi$, which are in the fundamental representation of the $SU(N)_T$ gauge group and are also charged under $SU(2)_W\times U(1)_Y$ gauge group;
 $\chi$ is an isospin doublet with hypercharge $Y=b/2$ and $\psi$ is an isospin singlet with hypercharge $Y=(b-1)/2$, with $b$ being an arbitrary number.
There also is a massless elementary scalar field, $H$, which possesses the same electroweak charge as the SM Higgs field and is a singlet under the $SU(N)_T$ gauge group.
The charges of the fields $\chi,\psi$ and $H$ are summarized in Table~\ref{fields}.
\begin{table}[h]
\begin{center}
\begin{tabular}{|c|c|c|c|c|} \hline
Field      & Lorentz $SO(1,3)$ & $SU(N)_T$ & $SU(2)_W$ & $U(1)_Y$ \\ \hline
$H$        & \bf{1}              & \bf{1}       & \bf{2}   & $1/2$ \\
$\chi$     & \bf{(2, 2)}         & \bf{N}       & \bf{2}   & $b/2$ \\
$\psi$     & \bf{(2, 2)}         & \bf{N}       & \bf{1}   & $(b-1)/2$ \\ \hline
\end{tabular}
\end{center}
\caption{The charges of the fields $\chi,\psi$ and $H$.
Representations in the Lorentz group, the $SU(N)_T$ gauge group and the SM electroweak gauge groups $SU(2)_W\times U(1)_Y$ are displayed.
$b$ is an arbitrary number.}
\label{fields}
\end{table}
Classical scale invariance forbids Dirac mass terms for $\chi$ and $\psi$ and a mass term for $H$,
 while $\chi$ and $\psi$ are allowed to have a Yukawa-type coupling with $H$.
We further assume invariance of the theory under the parity transformation on $\chi$ and $\psi$, $\chi \to \gamma_0\chi$, and $\psi \to \gamma_0\psi$.
This assumption is required to fully use analogy with QCD.
The Lagrangian for $\chi,\psi$ and $H$ fields thus reads
\begin{align}
{\cal L}_{{\rm ewsb}} &= \left\vert \left( \partial_{\mu} + i g_W W^a_{\mu} \tau^a + i \frac{1}{2} g_Y B_{\mu} \right) H \right\vert^2
\nonumber \\
&+ i\bar{\chi} \gamma^{\mu} \left( \partial_{\mu} + i g_T X_{\mu}^\alpha T^\alpha + i g_W W_{\mu}^a \tau^a + i \frac{b}{2} g_Y B_{\mu} \right) \chi 
+ i\bar{\psi} \gamma^{\mu} \left( \partial_{\mu} + i g_T X_{\mu}^\alpha T^\alpha + i \frac{b-1}{2} g_Y B_{\mu} \right) \psi
\nonumber \\
&- y \, \bar{\chi} \psi H - y \, H^{\dagger} \bar{\psi} \chi
\label{lagrangian1} \\
&- \lambda \, (H^{\dagger}H)^2 -Y_u \, \bar{q}u \, \epsilon H^* - Y_d \, \bar{q}d \, H - Y_e \, \bar{\ell}e \, H - {\rm H.c.},
\label{lagrangian2}
\end{align}
 where $y$ is a Yukawa coupling constant for $H$, $\chi$ and $\psi$ fields, which is taken to be real by phase redefinition of $\psi$ field.
Here, $g_T$, $g_W$ and $g_Y$ denote the $SU(N)_T$, SM weak and hypercharge gauge coupling constants,
 $X_{\mu}^\alpha$, $W_{\mu}^a$ and $B_{\mu}$ denote the corresponding gauge fields,
 and $T^\alpha \, (\alpha=1,2,...,N^2-1)$ and $\tau^a \, (a=1,2,3)$ are generators of the $SU(N)_T$ and weak gauge groups, respectively.
$q$, $u$, $d$, $\ell$ and $e$ respectively denote SM isospin doublet quarks, singlet up-type quarks, down-type quarks, doublet leptons and singlet leptons,
 $Y_u$, $Y_d$ and $Y_e$ are coupling constants proportional to the SM Yukawa couplings (flavor indices are omitted), and $\epsilon$ denotes the antisymmetric tensor in the isospin space.

The $SU(N)_T$ gauge theory is assumed to become strongly-coupled at infrared scales and induce chiral symmetry breaking and confinement of $\chi$,$\psi$ fields in the same way as QCD.
We infer the pattern of chiral symmetry breaking from the most attractive channel hypothesis.
When applied to weak and hypercharge gauge boson exchange forces~\cite{mac}, the hypothesis argues that
 composite operators with the smallest values of the quadratic Casimir operators of $SU(2)_W$ and $U(1)_Y$ groups form condensates,
 while when applied to the exchange of the elementary scalar $H$~\cite{scalarmac}, those with the largest values of the quadratic Casimir operators are expected to form condensates.
In the current model, we assume $y$ to be sufficiently small that the $\bar{\chi}\psi H$ Yukawa interaction is subdominant compared to the weak and hypercharge gauge interactions,
 which gives that operators with the smallest quadratic Casimir operators go into chiral condensation, namely, we have
\begin{align}
\langle 0 \vert \bar{\chi}\chi \vert 0 \rangle &\neq 0, \ \ \ \ \ \langle 0 \vert \bar{\chi}\tau^a\chi \vert 0 \rangle = 0 \ (a=1,2,3), \ \ \ \ \ \langle 0 \vert \bar{\psi}\psi \vert 0 \rangle \neq 0, \ \ \ \ \ \langle 0 \vert \bar{\chi}\psi \vert 0 \rangle = 0.
\label{chsb}
\end{align}
Note that the electroweak symmetry is maintained at this stage, unlike in the technicolor model~\cite{tc}.
This owes to the fact that $\chi,\psi$ fields are vector-like with respect to the weak and hypercharge gauge groups.

In the mass spectrum below the confinement scale, there exists a scalar meson, $\Theta$, that corresponds to a scalar bound state of $\bar{\psi}\chi$.
We write the mass of $\Theta$ meson as $M_\Theta$, and further define the scalar decay constant for $\Theta$ meson, $F_\Theta$, in the following fashion:
\begin{align}
\hat{y} F_\Theta M_\Theta &\equiv \langle 0 \vert y \, \bar{\psi}\chi(0) \vert \Theta \rangle,
\label{decayconstant}
\end{align}
 where $\vert \Theta \rangle$ denotes the state with one $\Theta$ meson, $y \, \bar{\psi}\chi$ denotes a scalar current accompanied by the coupling constant $y$,
 and $\hat{y}$ is a quantity proportional to $y$ that is RG-invariant in the $SU(N)_T$ gauge theory.
Note that inclusion of the coupling constant $y$ in the definition of the scalar current is advantageous in that the current becomes independent of the wavefunction renormalization in the $SU(N)_T$ gauge theory.
Below the confinement scale, $\bar{\psi}\chi$ term in the Lagrangian~Eq.~(\ref{lagrangian1}) asymptotes to $\Theta$ meson term,
 and accordingly, $y \, H \bar{\chi}\psi$ term becomes a mixing term for $H$ and $\Theta$ as
\begin{align}
y \, \bar{\chi}\psi \, H \ &\Rightarrow \ (\hat{y} F_\Theta M_\Theta) \, \Theta^{\dagger}H.
\end{align}
The mass matrix for $H$ and $\Theta$ is thus found to be
\begin{align}
-{\cal L}_{{\rm ewsb}} &\supset \left(
\begin{array}{cc}
H^{\dagger} & \Theta^{\dagger}
\end{array}
\right)
\left(
\begin{array}{cc}
0 & \hat{y} F_\Theta M_\Theta \\
\hat{y} F_\Theta M_\Theta & M_\Theta^2
\end{array}
\right)
\left(
\begin{array}{c}
H  \\
\Theta
\end{array}
\right).
\label{hthetamass}
\end{align}
We hereafter concentrate on the limit with small $\hat{y}$ that leads to $\hat{y} F_\Theta \ll M_\Theta$.
Upon diagonalization, we obtain the following mass terms for mass eigenstate scalar fields $H_1$ and $H_2$:
\begin{align}
-{\cal L}_{{\rm ewsb}} &\supset -(\hat{y} F_\Theta)^2 \, H_1^\dagger H_1 + M_\Theta^2 \, H_2^\dagger H_2,
\nonumber \\
&\left(
\begin{array}{c}
H  \\
\Theta
\end{array}
\right)=
\left(
\begin{array}{cc}
c_H & s_H \\
-s_H & c_H
\end{array}
\right)
\left(
\begin{array}{c}
H_1  \\
H_2
\end{array}
\right), \ \ \ s_H = \frac{\hat{y}F_\Theta}{M_\Theta}, \ 
c_H =  1 - \frac{1}{2} \frac{(\hat{y}F_\Theta)^2}{M_\Theta^2}.
\label{diag}
\end{align}
Since $H_1$ has a negative mass squared term, it develops a non-zero vacuum expectation value (VEV) and triggers electroweak symmetry breaking.
Hence, we identify $H_1$ with the SM Higgs field.
The $H_1$ field has a quartic coupling and Yukawa couplings induced from Eq.~(\ref{lagrangian2}) as
\begin{align}
-{\cal L}_{{\rm ewsb}} &\supset \lambda \, c_H^4 \, (H_1^\dagger H_1)^2 + Y_u \, c_H \, \bar{q}u \, \epsilon H_1^* + Y_d \, c_H \, \bar{q}d \, H_1 + Y_e \, c_H \, \bar{\ell}e \, H_1 + {\rm H.c.}
\nonumber \\ &{\rm with \ } c_H =  1 - \frac{1}{2} \frac{(\hat{y}F_\Theta)^2}{M_\Theta^2}.
\end{align}
The mass term, quartic coupling and Yukawa couplings of $H_1$ field should agree with those of the SM Higgs field at the energy scale of the $H_2$ mass, that is, $M_\Theta$.
These requirements are encapsulated in the following matching conditions:
\begin{align}
-\frac{1}{2}m_h^2 = ({\rm SM \ Higgs \ field \ mass}) &= -(\hat{y}F_\Theta)^2,
\label{matching1}
\\
\lambda^{{\rm SM}}(M_\Theta) &= \lambda(M_\Theta) \, c_H^4 = \lambda(M_\Theta) \, \left(1-2\frac{(\hat{y}F_\Theta)^2}{M_\Theta^2} \right),
\label{matching2}
\\
Y_k^{{\rm SM}}(M_\Theta) &= Y_k(M_\Theta) \, c_H = Y_k(M_\Theta) \, \left(1-\frac{1}{2}\frac{(\hat{y}F_\Theta)^2}{M_\Theta^2} \right) \ \ \ (k=u,d,e), \nonumber \\
\label{matching3}
\end{align}
 where $m_h$ denotes the pole mass of the SM Higgs particle, and $\lambda^{{\rm SM}}(M_\Theta)$ and $Y_u^{{\rm SM}}(M_\Theta)$, $Y_d^{{\rm SM}}(M_\Theta)$, $Y_e^{{\rm SM}}(M_\Theta)$ respectively represent the SM Higgs quartic coupling and SM Yukawa couplings
 at the energy scale $M_\Theta$.

In this way, the Higgs field mass is generated at the scale determined by the $\Theta$ meson decay constant $F_\Theta$ and the $\bar{\chi}\psi H$ Yukawa coupling constant $y$,
 with the correct negative sign owing to the negative determinant of the mass matrix Eq.~(\ref{hthetamass}).
The scale and sign of the SM Higgs mass term are thus attributed to strong dynamics of the $SU(N)_T$ gauge theory and its coupling with the elementary scalar field.
\\

We comment in passing that the elementary scalar $H$ does not have a mixing term with
any pNG boson associated with the chiral symmetry breaking, due to the assumption that
the theory is invariant under the parity transformation on $\chi$ and $\psi$ that makes the Yukawa
coupling appear in the form Eq.~(\ref{lagrangian1}). To prove this, we employ chiral perturbation theory: First
we define the fields,
\begin{align} 
\Psi_L &\equiv \frac{1-\gamma_5}{2}
\left(
\begin{array}{c}
\chi  \\
\psi
\end{array}
\right),
\ \ \
\Psi_R \equiv \frac{1+\gamma_5}{2}
\left(
\begin{array}{c}
\chi  \\
\psi
\end{array}
\right),
\end{align}
in terms of which the Yukawa interaction of Eq.~(\ref{lagrangian1}) is expressed as
\footnote{
We here refrain from the phase redefinition of $H$ that makes $y$ real.
}
\begin{align} 
-{\cal L} &\supset \bar{\Psi}_L
\left(
\begin{array}{cc}
0 & yH  \\
y^*H^\dagger & 0
\end{array}
\right)
\Psi_R
+ {\rm h.c.}
\end{align}
The $SU(N)_T$ gauge theory possesses approximate $U(3)_L \times U(3)_R$ symmetry under the transformation
$\Psi_L \to U_L \Psi_L$, $\Psi_R \to U_R \Psi_R$, with $U_L$, $U_R$ being unitary matrices.
Along the chiral symmetry breaking, the axial part $U(3)_A$ is spontaneously broken and there appear 9~pNG
bosons (one of which gains mass at the dynamical scale from instantons). We parametrize
them as
\begin{align} 
U(x) &= \exp\left( \ 2i \frac{\Pi^j(x)}{f_\Pi}\frac{\lambda^j}{2} + \sqrt{2}i \frac{\Pi(x)}{f_\Pi} \ \right),
\end{align}
where $\Pi^j(x)$ $(j=1,2,...,8)$ and $\Pi(x)$ denote the pNG boson fields, $\lambda^j$'s are the Gell-Mann
matrices, and $f_\Pi$ is the NG boson decay constant. $U(x)$ transforms under $U(3)_L \times U(3)_R$ as
\begin{align} 
U(x) &\to U_R U(x) U_L^\dagger.
\end{align}
Since we can assign to the Yukawa coupling and $H$ the following spurious transformation property,
\begin{align} 
\left(
\begin{array}{cc}
0 & yH  \\
y^*H^\dagger & 0
\end{array}
\right)
&\to
U_L
\left(
\begin{array}{cc}
0 & yH  \\
y^*H^\dagger & 0
\end{array}
\right)
U_R^\dagger,
\end{align}
the effective Lagrangian reads
\begin{align} 
{\cal L}_{eff} &= \frac{f_\Pi^2}{4} {\rm tr}\left[ \left(D_\mu U(x)\right)^\dagger D^\mu U(x) \right] 
- B_0 \ {\rm tr}\left[U(x)
\left(
\begin{array}{cc}
0 & yH  \\
y^*H^\dagger & 0
\end{array}
\right)
\right]
- {\rm h.c.},
\label{efflag}
\end{align}
where $B_0$ is a constant and $D_\mu$ is a covariant derivative. Based on analogy with QCD, we
expect $B_0$ to be real. Then the second term of Eq.~(\ref{efflag}) is recast into the form,
\begin{align} 
-{\cal L}_{eff} &\supset B_0 \ {\rm tr}\left[ \ \left\{ U(x)+U(x)^\dagger \right\} \ 
\left(
\begin{array}{cc}
0 & yH  \\
y^*H^\dagger & 0
\end{array}
\right)
\right].
\end{align}
Since the above interaction term is symmetric under the transformation $\Pi^j \to -\Pi^j$ or $\Pi \to -\Pi$,
the couplings between $H$ and the pNG bosons contain even numbers of pNG bosons and hence
no mixing term exists between them.
\\

\subsection{Evaluation of the dynamical scale of the $SU(N)_T$ gauge theory}

We evaluate the dynamical scale of the $SU(N)_T$ gauge theory from the matching condition~Eq.~(\ref{matching1})
 under analogy between the $SU(N)_T$ gauge theory and QCD.
This is done by expressing the $\Theta$ meson mass, total width and decay constant
 in terms of corresponding quantities in QCD, rescaled by 
 the ratio of the dynamical scales of the $SU(N)_T$ gauge theory and QCD, $r$, defined by
\begin{align}
r &\equiv \frac{\Lambda_T}{\Lambda_{QCD}},
\end{align}
 where $\Lambda_T$ and $\Lambda_{QCD}$ denote the dynamical scales of the $SU(N)_T$ gauge theory and QCD, respectively.

Along with $r$, a factor depending on $N/N_c$ ($N_c=3$ is the number of colors in QCD) appears when one relates some quantities in the $SU(N)_T$ gauge theory to their QCD counterparts.
In this paper, we make the two approximations below:
\\

\noindent
($\alpha$) The Casimir operator $C_F$ is approximated as $C_F=(N^2-1)/(2N)\simeq N/2$. 

\noindent
($\beta$) The contribution of $\chi,\psi$ fermion loop to the gauge field propagator is subdued compared to those of gauge field loop and ghost loop, and is hence negligible.
For example, the one-loop correction to the gauge field propagator is proportional to $11N-2n_f$ with $n_f$ being the number of flavors which equals $3$ in our model,
 and this is approximated as $11N-2n_f=11N-6\simeq 11N$.
\\

\noindent
($\alpha$) and ($\beta$) give that the coefficient for $O(\alpha_T^n)$ correction term is proportional to $N^n$.
It immediately follows that the $SU(N)_T$ gauge coupling $g_T$ and the QCD gauge coupling $g_s$ satisfy $g_T^2(r\mu)\simeq(N_c/N)g_s^2(\mu)$.
Also, correlation functions of $SU(N)_T$-singlet operators, with all operators connected, are obtained by
 rescaling corresponding correlation functions in QCD by $N/N_c$, \textit{e.g.}, we have
\begin{align}
\langle 0 \vert T\{ \bar{f}\Gamma_1 f(x_1) \, \bar{f}\Gamma_2 f(x_2) \, ... \, \bar{f}_n\Gamma_n f_n(x_n) \} \vert 0 \rangle &=
\frac{N}{N_c} r^{3n} \, \langle 0 \vert T\{ \bar{q}\Gamma_1 q(x_1) \, \bar{q}\Gamma_2 q(x_2) \, ... \, \bar{q}\Gamma_n q(x_n) \} \vert 0 \rangle,
\label{scaling}
\end{align}
 where $\bar{f}\Gamma_i f$ and $\bar{q} \Gamma_i q$ ($i=1,2,...,n$) represent $SU(N)_T$-singlet and QCD-singlet bilinear operators of massless Dirac fermions 
 in the fundamental representation of the $SU(N)_T$ and QCD gauge groups, respectively, with $\Gamma_i$ being a combination of the gamma matrices.
As a corollary, the wavefunction overlap of a one-meson-state with a $SU(N)_T$-singlet current operator
 roughly scales by $\sqrt{N/N_c}$, because the two-point self-correlation function of that current operator scales by $N/N_c$.
In particular, the decay constant for a Nambu-Goldstone (NG) boson is proportional to $\sqrt{N/N_c}$.
Moreover, the decay amplitude for a meson decaying into two mesons through the $SU(N)_T$ gauge interaction scales by $(N/N_c)^{-1/2}$.
This is because a correlation function for three $SU(N)_T$-singlet current operators scales by $N/N_c$,
 and when the current operators asymptote to one-meson creation operators, a factor $(N/N_c)^{-1/2}$ appears for each current,
 which leads to an overall factor of $(N/N_c)\{(N/N_c)^{-1/2}\}^3=(N/N_c)^{-1/2}$ for the decay amplitude.
\\

\subsubsection{$\Theta$ meson mass $M_\Theta$}

We evaluate the $\Theta$ meson mass $M_\Theta$ by regarding the $K_0^*(1430)$ meson as a QCD analog of the $\Theta$ meson for two reasons
\footnote{
For studies on scalar mesons, see Ref.~\cite{scalarmesons} and references therein.
}:
\\

\noindent
 (i)It is likely that $K_0^*(1430)$ belongs to a nonet of quark-anti-quark bound states whereas $K_0^*(800)$ belongs a nonet of diquark-anti-diquark bound states,
 notably because the $K_0^*(800)$ mass is considerably smaller than the $a_0(980)$ mass, which suggests that $K_0^*(800)$ is mainly a bound state of $(s,q,\bar{q},\bar{q}')$ and
 $a_0(980)$ is that of $(s,q,\bar{s},\bar{q}')$ ($q,q'$ denote up and down quarks and $s$ denotes strange quark).
 Hence, $K_0^*(1430)$ meson corresponds to the lightest $\bar{\chi}\psi$ scalar bound state, \textit{i.e.}, $\Theta$ meson.

\noindent
 (ii)$K_0^*(1430)$ meson and $\Theta$ meson do not mix with glueball state and its $SU(N)_T$-gauge-theory counterpart due to non-zero strangeness and electroweak charge, respectively.
\\

\noindent
Analogy between $\Theta$ meson and $K_0^*(1430)$ meson enables us to express the $\Theta$ meson mass $M_\Theta$ in terms of the $K_0^*(1430)$ mass, $m_{K_0^*(1430)}$, as
\begin{align}
M_\Theta &= r \, m_{K_0^*(1430)}.
\label{mtheta}
\end{align}
The experimental central value~\cite{pdg} gives $M_\Theta = r\cdot 1.425$~GeV.
\\

\subsubsection{$\Theta$ meson total width $\Gamma_\Theta$}

The $\Theta$ meson total width, $\Gamma_\Theta$, can be written with the $K_0^*(1430)$ meson total width, $\Gamma_{K_0^*(1430)}$.
In doing so, we remind that about 90\% of $K_0^*(1430)$ decays as $K_0^*(1430) \to K \pi$~\cite{pdg}, in which process the $K,\pi$ meson masses modify the phase space,
 while the corresponding pNG bosons in the $SU(N)_T$ gauge theory are nearly massless due to zero current mass.
Hence, we rescale the $K_0^*(1430)$ total width by the phase space ratio, together with the dynamical scale ratio $r$ and the factor $(N/N_c)^{-1}$, to evaluate $\Gamma_\Theta$.
It is thus found to be
\begin{align}
\Gamma_\Theta &= r \, \frac{N_c}{N} \, \frac{1}{ \sqrt{1-2\frac{m_\pi^2+m_K^2}{m_{K_0^*(1430)}^2}+\frac{(m_\pi^2-m_K^2)^2}{m_{K_0^*(1430)}^4}} } \, \Gamma_{K_0^*(1430)},
\label{gammatheta}
\end{align}
 where $m_\pi$ and $m_K$ denote the $\pi$ and $K$ masses, respectively.
The experimental central values~\cite{pdg} give $\Gamma_\Theta = r(N_c/N)\cdot 0.311$~GeV.
\\

\subsubsection{$\Theta$ meson scalar decay constant $F_\Theta$}

Since a QCD analog of the $\Theta$ meson scalar decay constant $F_\Theta=\langle 0 \vert y \, \bar{\psi}\chi \vert \Theta \rangle/(\hat{y}M_\Theta)$ has not been measured,
 we evaluate it by the aid of explicit calculations.
To this end, we confront the spectral density of states that couple to the scalar current $y \, \bar{\chi}\psi$, with the two-point self-correlation function of that scalar current for large space-like momenta;
the former spectral density contains the term $\hat{y}F_\Theta M_\Theta$, as it quantifies the coupling of the one-$\Theta$-meson state with the scalar current.
The latter correlation function can be described with perturbative calculation in the $SU(N)_T$ gauge theory and with empirical vacuum condensates.
To embody the idea, we formulate the correlation function, $\Pi_{y\bar{\chi}\psi}(q^2)$, and the spectral density, $\rho_s(s)$, in the following fashion
\footnote{
We insert the factor $1/2$ in Eq.~(\ref{correlatordef}) to cancel the $SU(2)_W$ degree of freedom of $\chi$.
}:
\begin{align}
\Pi_{y\bar{\chi}\psi}(q^2) &\equiv \frac{1}{2}i\int{\rm d}^4x \, e^{iqx} \langle 0 \vert T\left\{ y \, \bar{\chi}(x)\psi(x) \ y \, \bar{\psi}(0)\chi(0) \right\} \vert 0 \rangle,
\label{correlatordef} \\
\rho_s(s) &\equiv {\rm Im}\Pi_{y\bar{\chi}\psi}(s) \ \ \ \ \ {\rm for} \ s\geq0.
\label{spectraldef}
\end{align}
We relate the spectral density to the correlation function for space-like momenta using the Cauchy's theorem.
Since no one-massless-particle state couples to the scalar current, we have
\begin{align}
\lim_{\vert q^2\vert\to0} \, q^2\Pi_{y\bar{\chi}\psi}(q^2) &= 0
\end{align}
 for complex $q^2$, which allows us to use the Cauchy's theorem to obtain
\begin{align}
\int_0^{s_0}{\rm d}s \, \{P(s)-P(s_0)\} \, \frac{1}{\pi}{\rm Im}\Pi_{y\bar{\chi}\psi}(s) = -\frac{1}{2\pi i}\oint_{\vert q^2 \vert=s_0}{\rm d}q^2 \, \{P(q^2)-P(s_0)\} \, \Pi_{y\bar{\chi}\psi}(q^2),
\label{cauchy}
\end{align}
 where $s_0$ can take any real positive value and $P(s)$ can be any analytic function.
Here, on the right-hand side, the correlation function for complex $q^2$ off the real positive axis can be derived through analytic continuation from space-like momenta,
 while the discontinuity of the correlation function at $q^2=s_0$ is avoided as the function $P(q^2)-P(s_0)$ vanishes there.
The left-hand side includes $\hat{y}F_\Theta M_\Theta$,
 and the right-hand side is calculable by perturbation theory and operator product expansion in the $SU(N)_T$ gauge theory if $s_0$ is taken sufficiently large that these calculations are valid.
Therefore, the equality Eq.~(\ref{cauchy}) enables us to evaluate $\hat{y}F_\Theta M_\Theta$ through analytic calculations.
The use of Eq.~(\ref{cauchy}) for such $s_0$ constitutes the basis for the finite energy sum rules~\cite{fesr}.

First we express the spectral density $\rho_s(s)$ in terms of $\hat{y}F_\Theta M_\Theta$ based on the following two assumptions on quantum states that couple to the scalar current $y \, \bar{\chi}\psi$:
\\

\noindent
(i) In the range $M_\Theta^2-r^2\cdot1~{\rm GeV}^2 \lesssim s \lesssim M_\Theta^2+r^2\cdot1~{\rm GeV}^2$, the spectral density $\rho_s(s)$ is dominated by contributions from the $\Theta$ meson resonance.

\noindent
(ii) The $\Theta$ meson resonance is well approximated by the relativistic Breit-Wigner function.
\\

\noindent
These ansaetze find no support from hadron physics experiments, even given analogy between $\Theta$ meson and $K_0^*(1430)$ meson, because an elementary scalar current that couples to light quarks has not been measured.
Nevertheless, $S$-wave $K\pi$ scattering data from the LASS experiment~\cite{lass} hint us that the $K_0^*(1430)$ meson resonance may dominantly couple to 
 $\bar{d}s$ scalar current ($d$ and $s$ respectively denote down and strange quarks) for invariant mass $m_{K_0^*(1430)}^2-1~{\rm GeV}^2 \lesssim s \lesssim m_{K_0^*(1430)}^2+1~{\rm GeV}^2$,
 which is rendered into the assumption (i) through the rescaling.
The same data also suggest that the $K_0^*(1430)$ meson resonance can be fit with the relativistic Breit-Wigner function.
Once one accepts (i) and (ii), the spectral density satisfies
\begin{align}
{\rm for} \ &M_\Theta^2-r^2\cdot1~{\rm GeV}^2 \lesssim s \lesssim M_\Theta^2+r^2\cdot1~{\rm GeV}^2,
\nonumber \\
\rho_s(s) &= \frac{1}{\pi}{\rm Im}\Pi_{y\bar{\chi}\psi}(s)
\nonumber \\
&= \frac{1}{\pi}{\rm Im}\left[ \, \langle 0 \vert y \, \bar{\psi}\chi \vert \Theta \rangle \frac{1}{s-M_\Theta^2+iM_\Theta\Gamma_\Theta} \langle \Theta \vert y \, \bar{\chi}\psi \vert 0 \rangle \, \right]
\nonumber \\
&= (\hat{y} F_\Theta M_\Theta)^2 \, \frac{1}{\pi}\frac{M_\Theta \Gamma_\Theta}{(s-M_\Theta^2)^2+M_\Theta^2 \Gamma_\Theta^2}.
\label{spectral}
\end{align}
Note here that since $\Theta$ meson decays into two pNG bosons, which are nearly massless due to the absence of current mass, the Breit-Wigner function reduces to the form as appears in Eq.~(\ref{spectral}).

Next we present the scalar current correlation function for space-like momenta $q^2 < 0$ as a perturbative series of the $SU(N)_T$ gauge coupling and an operator product expansion involving vacuum condensates,
 following analogous calculations for QCD in Refs.~\cite{chetyrkin, chetyrkin2, qcdsumrules, bagan}.
The scalar current correlation function is given, for $q^2 < 0$, by
\begin{align}
&\Pi_{y\bar{\chi}\psi}(q^2) = A \, q^2 + y^2(\mu^2) \frac{N}{8\pi^2}(-q^2) \log\left(\frac{-q^2}{\mu^2}\right) \left\{ \, 1
+ \sum_{k=1}^\infty \, \left(\frac{\alpha_T(\mu^2)}{\pi}\right)^k \, \sum_{l=0}^k \, d_{k,l}(C_F,C_A) \, \log^l\left(\frac{-q^2}{\mu^2}\right) \, \right\}
\nonumber \\
&+ \frac{1}{8\pi}\frac{1}{(-q^2)} \, y^2(\mu^2) \, \langle 0 \vert \alpha_T X_{\mu\nu}^\alpha X^{\alpha\mu\nu} \vert 0 \rangle
\nonumber \\
&+ 2\pi\frac{1}{(-q^2)^2} y^2(\mu^2) \langle 0 \vert \alpha_T \{ \, \bar{\chi}\sigma_{\mu\nu} T^\alpha \psi \, \bar{\psi}\sigma^{\mu\nu} T^\alpha \chi + \frac{1}{3}(\bar{\chi}\gamma_\mu T^\alpha \chi+2\bar{\psi}\gamma_\mu T^\alpha \psi)(\bar{\chi}\gamma^\mu T^\alpha \chi+\bar{\psi}\gamma^\mu T^\alpha \psi) \, \} \vert 0 \rangle
\nonumber \\
&+ ({\rm other \ vacuum \ condensates}) + ( \ O(\alpha_T(\mu^2)/\pi){\rm \ corrections \ to \ vacuum \ condensates} \ ).
\label{correlator}
\end{align}
Here, $A$ is a constant of no interest, $\alpha_T\equiv g_T^2/(4\pi)$, $C_F=(N^2-1)/(2N)$, $C_A=N$, $T^\alpha$ denotes $SU(N)_T$ group generators, and $X_{\mu\nu}^\alpha$ denotes the $SU(N)_T$ gauge field strength.
$\mu$ denotes the renormalization scale in the $\overline{MS}$ scheme for the correlation function itself, the running coupling constant $y(\mu^2)$ and the running gauge coupling constant $\alpha_T(\mu^2)$, which is taken to be common for simplicity.
$d_{k,l}(C_F,C_A)$'s are coefficients that depend on $C_F,C_A$, whose explicit forms are found in Ref.~\cite{chetyrkin}.
The first term in the last line represents vacuum condensates of other operators, which are known to be subdominant compared to the field strength condensate and four-fermion condensate for $-q^2 \gtrsim r^2\cdot1$~GeV$^2$.
We stress that fermion bilinear condensates $\langle 0 \vert \bar{f}f \vert 0 \rangle$ ($f=\chi,\psi$) do not enter into Eq.~(\ref{correlator}) due to the absence of $\chi,\psi$ current mass.

\begin{figure}[H]
  \begin{center}
    \includegraphics[width=120mm]{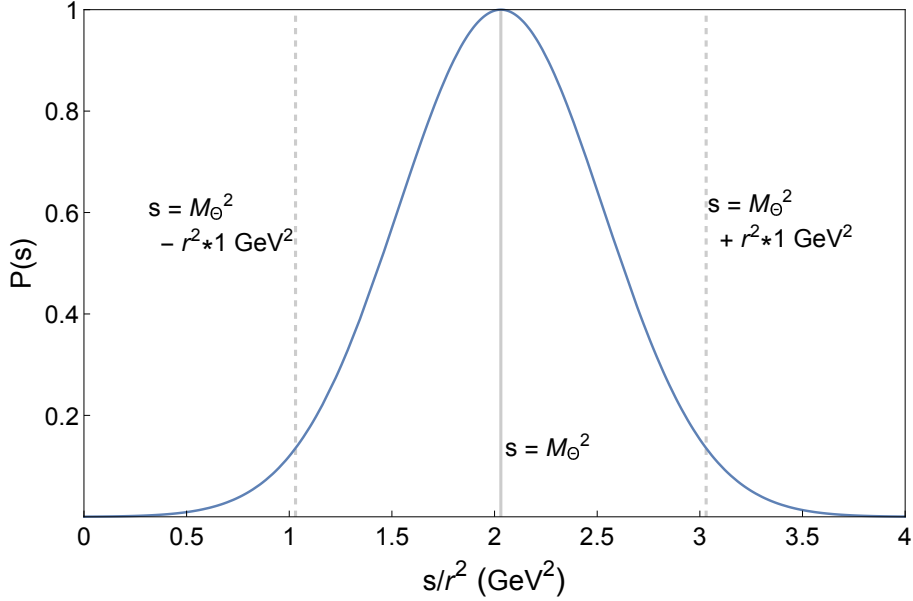}
    \caption{
    The function $P(s)$ Eq.~(\ref{windowfunc}) in the range $0\leq s \leq r^2\cdot4~{\rm GeV}^2$.
    The solid vertical line corresponds to $s=M_\Theta^2$ and dashed lines to $s=M_\Theta^2\pm r^2\cdot1~{\rm GeV}^2$.
    }
    \label{pfunc}
  \end{center}
\end{figure}

Finally, we associate the spectral density Eq.~(\ref{spectral}) with the correlation function Eq.~(\ref{correlator}) through the Cauchy's theorem Eq.~(\ref{cauchy}).
For an arbitrary analytic function $P(s)$, we choose
\begin{align}
P(s) &= \exp\left[ \, -\frac{(s-M_\Theta^2)^2}{\Delta s^2} \, \right]
\ \ 
{\rm with \ \ } \Delta s = r^2 \cdot 1~{\rm GeV}^2,
\label{windowfunc}
\end{align}
 whose shape is depicted in Figure~\ref{pfunc}.
\noindent
The function $P(s)$ as given in Eq.~(\ref{windowfunc}) is beneficial for extracting the $\Theta$ meson contribution to the spectral density,
 because $P(s)$ is sizable only in the range $M_\Theta^2-r^2\cdot1~{\rm GeV}^2 \lesssim s \lesssim M_\Theta^2+r^2\cdot1~{\rm GeV}^2$, where the $\Theta$ meson resonance is assumed to dominate the spectral density,
 while contributions from the spectral density outside the range are exponentially suppressed.
We set the common renormalization scale for $\Pi_{y\bar{\chi}\psi}(q^2)$ as $\mu^2=s_0$.
We thus obtain
\begin{align}
&\int_0^{s_0}{\rm d}s \, \{P(s)-P(s_0)\} \, \rho_s(s) = -\frac{1}{2\pi i}\oint_{\vert q^2 \vert=s_0}{\rm d}q^2 \, \{P(q^2)-P(s_0)\} \, \Pi_{y\bar{\chi}\psi}(q^2)
\nonumber \\
&\Longleftrightarrow
\nonumber \\
&(\hat{y} F_\Theta M_\Theta)^2 \, \int_0^{s_0} {\rm d}s \, \{P(s)-P(s_0)\} \, \frac{1}{\pi} \frac{M_\Theta \Gamma_\Theta}{(s-M_\Theta^2)^2+M_\Theta^2 \Gamma_\Theta^2}
\nonumber \\
&= y^2(s_0) \frac{N}{8\pi^2} \, s_0^2 \, \left\{ \, B_0 + \sum_{k=1}^\infty \, \left(\frac{\alpha_T(s_0)}{\pi}\right)^k \, \sum_{l=0}^k \, d_{k,l}(C_F,C_A) \, B_{l+1} \, \right\}
\nonumber \\
&+ \frac{1}{8\pi} \, y^2(s_0) \, C_{1} \, \langle 0 \vert \alpha_T X_{\mu\nu}^\alpha X^{\alpha\mu\nu} \vert 0 \rangle
\nonumber \\
&+ 2\pi y^2(s_0) \, C_{2} \, \frac{1}{s_0} \langle 0 \vert \alpha_T \{ \, \bar{\chi}\sigma_{\mu\nu} T^\alpha \psi \, \bar{\psi}\sigma^{\mu\nu} T^\alpha \chi + \frac{1}{3}(\bar{\chi}\gamma_\mu T^\alpha \chi+2\bar{\psi}\gamma_\mu T^\alpha \psi)(\bar{\chi}\gamma^\mu T^\alpha \chi+\bar{\psi}\gamma^\mu T^\alpha \psi) \, \} \vert 0 \rangle
\nonumber \\
&+ ({\rm other \ vacuum \ condensates}) + ( \ O(\alpha_T(s_0)/\pi){\rm \ corrections \ to \ vacuum \ condensates} \ ),
\label{sumrule}
\end{align}
 where $B_l$'s and $C_{l}$'s are numbers defined as
\begin{align}
B_l &\equiv -\frac{1}{s_0^2} \, \frac{1}{2\pi i}\oint_{\vert q^2 \vert=s_0}{\rm d}q^2 \, \{ P(q^2)-P(s_0) \} \, (-q^2)\log^{l+1}\left(\frac{-q^2}{s_0}\right)
\nonumber \\
&= \frac{1}{2\pi i}\int_0^{2\pi}{\rm d}\theta \, i \, e^{2i\theta} \, \left\{ \, \exp\left[ \, -\frac{(s_0 e^{i\theta}-M_\Theta^2)^2}{2M^4} \, \right] - \exp\left[ \, -\frac{(s_0-M_\Theta^2)^2}{2M^4} \, \right] \, \right\} \, \{i(\theta-\pi)\}^{l+1},
\nonumber \\
C_{l} &\equiv -s_0^{l-1} \, \frac{1}{2\pi i}\oint_{\vert q^2 \vert=s_0}{\rm d}q^2 \, \{ P(q^2)-P(s_0) \} \, \frac{1}{(-q^2)^{l}}
\nonumber \\
&= \frac{(-1)^{l+1}}{2\pi i}\int_0^{2\pi}{\rm d}\theta \, i \, e^{-i \theta (l-1)} \, \left\{ \, \exp\left[ \, -\frac{(s_0 e^{i\theta}-M_\Theta^2)^2}{2M^4} \, \right] - \exp\left[ \, -\frac{(s_0-M_\Theta^2)^2}{2M^4} \, \right] \, \right\}.
\end{align}
One can derive $(\hat{y} F_\Theta M_\Theta)^2$ from the equality Eq.~(\ref{sumrule}).
The derived value would not depend on $s_0$ if the true spectral density were used and the correlation function
 were calculated to all orders in perturbation theory and in operator product expansion.
In reality, it does exhibit a $s_0$ dependence due to the error of spectral density assumed and the truncation of perturbative series and operator product expansion.
Nevertheless, we argue that a value of $(\hat{y} F_\Theta M_\Theta)^2$ that is most stable against variations of $s_0$
 is a physically meaningful estimate for $(\hat{y} F_\Theta M_\Theta)^2$.
To search for such a value, we numerically evaluate both sides of Eq.~(\ref{sumrule}) to express $(\hat{y} F_\Theta M_\Theta)^2$ as a function of $s_0$
\\

For the left-hand side of Eq.~(\ref{sumrule}), we numerically integrate the relativistic Breit-Wigner function Eq.~(\ref{spectral}) multiplied by the function $P(s)-P(s_0)$ Eq.~(\ref{windowfunc}),
 substituting the previously evaluated values of $\Theta$ meson mass $M_\Theta$ and total width $\Gamma_\Theta$.
Contributions from states other than the $\Theta$ meson resonance are ignored, as they are exponentially suppressed by the function $P(s)$.

In the right-hand side of Eq.~(\ref{sumrule}), we evaluate the perturbative series to the order of $\alpha_T^4$.
We quote the analytic expressions of $d_{1,l}$, $d_{2,l}$ and $d_{3,l}$ from Ref.~\cite{chetyrkin}.
As for $d_{4,0}$, only the formula for $N=3$ is available in the literature.
Therefore, we estimate $d_{4,0}$ for general $N$ by rescaling the analytic expression of $d_{4,0}$ for $N=3$ in Ref.~\cite{chetyrkin2} by the following rule, based on the approximation with $C_F=(N^2-1)/(2N)\simeq N/2$;
the $n_f^0$ term is rescaled by $(N/3)^4$; the $n_f^1$ term is rescaled by $(N/3)^3$; the $n_f^2$ term is rescaled by $(N/3)^2$; the $n_f^3$ term is rescaled by $N/3$, with $n_f$ denoting the number of quark flavors.
$d_{4,1},d_{4,2},d_{4,3},d_{4,4}$ are calculated from terms of order $\alpha_T^3$ or below through RG equations.
We separately evaluate $\alpha_T(s_0)$ and $y(s_0)$ by solving RG equations;
 to obtain $\alpha_T(s_0)$, we employ the $O(\alpha_s^4)$ RG equation for the QCD gauge coupling with the replacement of $C_F,C_A$ with those for the $SU(N)_T$ gauge group,
 and impose an initial condition at the scale $\mu=r\cdot1.777$~GeV. The equations are given below:
\begin{align}
\frac{{\rm d}\alpha_T(\mu^2)}{{\rm d}\log\mu^2} &= -\beta_1(C_F) \, \alpha_T^2(\mu^2) -\beta_2(C_F,C_A) \, \alpha_T^3(\mu^2) -\beta_3(C_F,C_A) \, \alpha_T^4(\mu^2),
\nonumber \\
\alpha_T(\mu^2) &= \frac{N_c}{N} \cdot 0.33 \ \ {\rm at} \ \ \mu=r\cdot1.777~{\rm GeV},
\end{align}
 where the forms of $\beta_1,\beta_2,\beta_3$ are found in Ref.~\cite{qcd3loop}.
Here, the initial condition is based on the QCD gauge coupling constant at the $\tau$ lepton mass scale $m_\tau = 1.777$~GeV studied in Ref.~\cite{pich},
 which is rescaled by the factor $N_c/N$ under the approximations $(\alpha),(\beta)$ in Section~2.2.
To gain $y(s_0)$, we exploit the $O(\alpha_s^4)$ RG equation for the running current quark mass in Ref.~\cite{larin} by exchanging $C_F,C_A$ with those for the $SU(N)_T$ gauge group.
This is justifiable because the coupling constant $y$ and the current quark mass obey the same RG equation in the $SU(N)_T$ gauge theory and QCD.
Since the overall scale of the coupling constant $y$ is a free parameter of the model, we express $y(s_0)$ as normalized by the value at one particular scale.
For later convenience, we express $y(s_0)$ in terms of $y$ evaluated at $\mu=r\cdot2$~GeV scale in the $\overline{MS}$ scheme, which we denote by $y_{r2}\equiv y(\mu^2=r^2\cdot2^2~{\rm GeV}^2)$.

The vacuum condensates are assessed by analogy with QCD;
 using the values of QCD gluon condensate and four-quark condensate obtained from $e^+e^-$ collisions, heavy quarkonia and $\tau$ lepton decays in Ref.~\cite{narison},
 we evaluate them as
\begin{align}
&\frac{1}{8\pi} \langle 0 \vert \alpha_T X_{\mu\nu}^\alpha X^{\alpha\mu\nu} \vert 0 \rangle
\simeq \frac{1}{8\pi} \, \frac{N}{N_c} \, r^4 \, \langle 0 \vert \alpha_s G_{\mu\nu}^aG^{a\mu\nu} \vert 0 \rangle = \frac{1}{8\pi} \, \frac{N}{N_c} \, r^4 \cdot 6.8\times10^{-2} \, {\rm GeV}^4,
\label{ggcondensate} \\
&2\pi\langle 0 \vert \alpha_T \{ \, \bar{\chi}\sigma_{\mu\nu}T^\alpha \psi \, \bar{\psi}\sigma^{\mu\nu}T^\alpha \chi + \frac{1}{3}(\bar{\chi}\gamma_\mu T^\alpha \chi + 2\bar{\psi}\gamma_\mu T^\alpha \psi)(\bar{\chi}\gamma^\mu T^\alpha \chi + \bar{\psi}\gamma^\mu T^\alpha \psi) \, \} \vert 0 \rangle 
\nonumber \\
&\simeq - \frac{22\pi}{3}\frac{N^2-1}{N^2} \, \rho \alpha_T(\langle 0 \vert \bar{f}f \vert 0 \rangle)^2 
\simeq -\frac{22\pi}{3}\frac{N^2-1}{N^2}\frac{N}{N_c} \, r^6 \, \rho\alpha_s(\langle 0 \vert \bar{q}q \vert 0 \rangle)^2 
\nonumber \\
&= -\frac{22\pi}{3}\frac{N^2-1}{N^2}\frac{N}{N_c} \, r^6 \cdot 4.5\times10^{-4} \, {\rm GeV}^6,
\label{4fcondensate}
\end{align}
 where in Eq.~(\ref{4fcondensate}), $f$ represents one massless Dirac fermion in the fundamental representation of the $SU(N)_T$ gauge group,
 and $\rho$ denotes the ratio of four-quark condensate and the square of quark bilinear condensate.
The factor $N/N_c$ ensues from the rescaling of the correlation functions of operators in the adjoint representation, which scale by $N^2/N_c^2$, 
 and that of the gauge coupling, which scales by $N_c/N$.

As a reference, we numerically express the perturbative series and the two vacuum condensates for $N=3,6$ and for $s_0=r^2\cdot3^2~{\rm GeV}^2, \, r^2\cdot5^2~{\rm GeV}^2$.
Also shown is the value of the gauge coupling $a\equiv (N/3)\alpha_T(s_0)/\pi$ for each $s_0$ and $N$.
We comment that for $s_0<r^2\cdot3^2~{\rm GeV}^2$, the perturbative series does not show good convergence, while for $s_0>r^2\cdot5^2~{\rm GeV}^2$, numerical calculation of the contour integral is computationally expensive due to a rapid oscillation of the function $P(q^2)$ for complex $q^2$.
\begin{align}
&{\rm For \ } N=3 {\rm \ and \ } s_0=r^2\cdot3^2~{\rm GeV}^2,
\nonumber \\
&-\frac{1}{2\pi i}\oint_{\vert q^2 \vert=s_0}{\rm d}q^2 \, \{P(q^2)-P(s_0)\} \, \Pi_{y\bar{\chi}\psi}(q^2)
\nonumber \\
&= y^2(s_0) \frac{N}{8\pi^2} \, s_0^2 \, ( \, 0.0444 + 0.378 \, a + 3.82 \, a^2 + 35.0 \, a^3  + 332 a^4 \, )
\nonumber \\
&+0.0162 \, \frac{y^2(s_0)}{8\pi} \frac{N}{3} \, r^4\cdot6.8\times10^{-2} \, {\rm GeV}^4 
+0.592 \, \frac{y^2(s_0)}{s_0}\frac{22\pi}{3}\frac{N^2-1}{N^2}\frac{N}{3} \, r^6\cdot4.5\times10^{-4} \, {\rm GeV}^6;
\nonumber \\
&a = \frac{N}{3}\frac{\alpha_T(s_0)}{\pi} = \frac{\alpha_T(s_0)}{\pi} = 0.0810.
\label{3r3} 
\end{align}
\begin{align}
&{\rm For \ } N=3 {\rm \ and \ } s_0=r^2\cdot5^2~{\rm GeV}^2,
\nonumber \\
&-\frac{1}{2\pi i}\oint_{\vert q^2 \vert=s_0}{\rm d}q^2 \, \{P(q^2)-P(s_0)\} \, \Pi_{y\bar{\chi}\psi}(q^2)
\nonumber \\
&= y^2(s_0) \frac{N}{8\pi^2} \, s_0^2 \, ( \, 0.00576 + 0.0608 \, a + 0.779 \, a^2 + 9.54 \, a^3 + 95.7 \, a^4 \, )
\nonumber \\
&+0.0162 \, \frac{y^2(s_0)}{8\pi} \frac{N}{3} \, r^4\cdot6.8\times10^{-2} \, {\rm GeV}^4 
+1.64 \, \frac{y^2(s_0)}{s_0}\frac{22\pi}{3}\frac{N^2-1}{N^2}\frac{N}{3} \, r^6\cdot4.5\times10^{-4} \, {\rm GeV}^6;
\nonumber \\
&a = \frac{N}{3}\frac{\alpha_T(s_0)}{\pi} = \frac{\alpha_T(s_0)}{\pi} = 0.0666.
\label{3r5}
\end{align}
\begin{align}
&{\rm For \ } N=6 {\rm \ and \ } s_0=r^2\cdot3^2~{\rm GeV}^2,
\nonumber \\
&-\frac{1}{2\pi i}\oint_{\vert q^2 \vert=s_0}{\rm d}q^2 \, \{P(q^2)-P(s_0)\} \, \Pi_{y\bar{\chi}\psi}(q^2)
\nonumber \\
&= y^2(s_0) \frac{N}{8\pi^2} \, s_0^2 \, ( \, 0.0444 + 0.414 \, a + 4.56 \, a^2 + 47.2 \, a^3 + 467 \, a^4 \, )
\nonumber \\
&+0.0162 \, \frac{y^2(s_0)}{8\pi} \frac{N}{3} \, r^4\cdot6.8\times10^{-2} \, {\rm GeV}^4 
+0.592 \, \frac{y^2(s_0)}{s_0}\frac{22\pi}{3}\frac{N^2-1}{N^2}\frac{N}{3} \, r^6\cdot4.5\times10^{-4} \, {\rm GeV}^6;
\nonumber \\
&a = \frac{N}{3}\frac{\alpha_T(s_0)}{\pi} = 2 \, \frac{\alpha_T(s_0)}{\pi} = 0.0784.
\label{6r3}
\end{align}
\begin{align}
&{\rm For \ } N=6 {\rm \ and \ } s_0=r^2\cdot5^2~{\rm GeV}^2,
\nonumber \\
&-\frac{1}{2\pi i}\oint_{\vert q^2 \vert=s_0}{\rm d}q^2 \, \{P(q^2)-P(s_0)\} \, \Pi_{y\bar{\chi}\psi}(q^2)
\nonumber \\ 
&= y^2(s_0) \frac{N}{8\pi^2} \, s_0^2 \, ( \, 0.00576 + 0.0665 \, a + 0.931 \, a^2 + 12.7 \, a^3  + 139 \, a^4 \, )
\nonumber \\
&+0.0162 \, \frac{y^2(s_0)}{8\pi}\frac{N}{3} \, r^4\cdot6.8\times10^{-2} \, {\rm GeV}^4 
+1.64 \, \frac{y^2(s_0)}{s_0}\frac{22\pi}{3}\frac{N^2-1}{N^2}\frac{N}{3} \, r^6\cdot4.5\times10^{-4} \, {\rm GeV}^6;
\nonumber \\
&a = \frac{N}{3}\frac{\alpha_T(s_0)}{\pi} = 2 \, \frac{\alpha_T(s_0)}{\pi} = 0.0634.
\label{6r5}
\end{align}
It is observed that the perturbative series converges at a rate about $0.7^k$ for all cases above.
We also find that for all $N$, the field-strength condensate $\langle 0 \vert \alpha_T X_{\mu\nu}^\alpha X^{\alpha\mu\nu} \vert 0 \rangle$ contributes to the right-hand side of Eq.~(\ref{sumrule}) by less than 0.03\% for $s_0=r^2\cdot3^2~{\rm GeV}^2$
 and by less than 0.004\% for $s_0=r^2\cdot5^2~{\rm GeV}^2$,
 while the four-fermion condensate contributes by less than 1\% for $s_0=r^2\cdot3^2~{\rm GeV}^2$ and by less than 0.05\% for $s_0=r^2\cdot5^2~{\rm GeV}^2$.
Since vacuum condensates have minor or negligible impact on the evaluation of $(\hat{y} F_\Theta M_\Theta)^2$,
 we ignore all vacuum condensates except the two displayed in Eq.~(\ref{sumrule}), and further discard $O(\alpha_T)$ corrections to vacuum condensates.

\begin{figure}[H]
  \begin{center}
    \includegraphics[width=100mm]{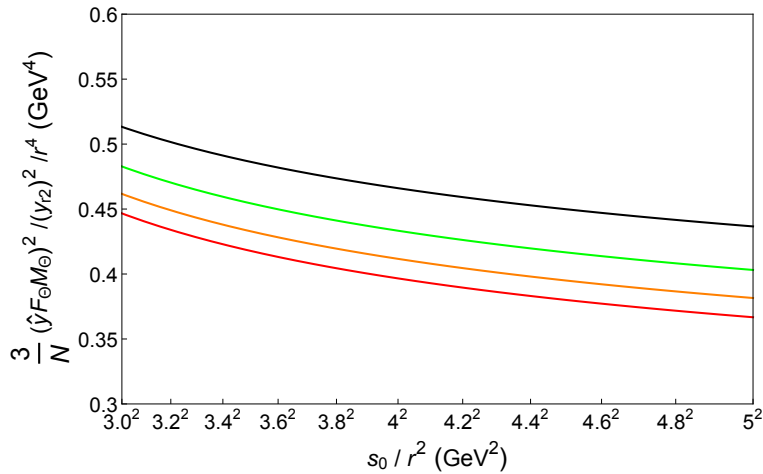}
    \caption{
    $(\hat{y} F_\Theta M_\Theta)^2$ as derived from Eq.~(\ref{sumrule}) with various values of $s_0$.
    The upper, middle-upper, middle-lower and lower lines respectively correspond to the cases with $N=3,4,5,6$.
    The value of $(\hat{y} F_\Theta M_\Theta)^2$ is normalized by $N/3$ and $y_{r2}$, which is the value of the coupling constant $y$ at $\mu=$2~GeV scale in the $\overline{MS}$ scheme.
    }
    \label{s0dep}
  \end{center}
\end{figure}

We plot in Figure~\ref{s0dep} the values of $(\hat{y} F_\Theta M_\Theta)^2$ derived from Eq.~(\ref{sumrule}) in the range $r^2\cdot3^2~{\rm GeV}^2 \leq s_0 \leq r^2\cdot5^2~{\rm GeV}^2$ for $N=3,4,5,6$.
We normalize $(\hat{y} F_\Theta M_\Theta)^2$ by $N/3$, as it is roughly linearly proportional to $N$.
It is further normalized by $y_{r2}$, the value of the coupling constant $y$ at $\mu=2$~GeV scale in the $\overline{MS}$ scheme.
\noindent
The decrease of $(\hat{y} F_\Theta M_\Theta)^2$ with $s_0$ is totally ascribed to the truncation of perturbative series at order $\alpha_T^4$,
 because the integral on the left-hand side of Eq.~(\ref{sumrule}), \textit{i.e.}, the integral of the Breit-Wigner function times the function $P(s)-P(s_0)$, is virtually constant for $s_0 \geq r^2\cdot3^2~{\rm GeV}^2$ due to exponential suppression in $P(s)$.
Since the variation of $(\hat{y} F_\Theta M_\Theta)^2$ with $s_0$, namely, steepness of the curves in Figure~\ref{s0dep}, falls off as $s_0$ approaches to $r^2\cdot5^2~{\rm GeV}^2$,
 we consider that the value corresponding to $s_0 = r^2\cdot5^2~{\rm GeV}^2$ is most close to the true physical value.
Hence, we conclude with the following estimate for $(\hat{y} F_\Theta M_\Theta)^2$ derived by setting $s_0 = r^2\cdot5^2~{\rm GeV}^2$ in Eq.~(\ref{sumrule}):
\begin{align}
(\hat{y} F_\Theta M_\Theta)^2 &= \frac{N}{3} \, y_{r2}^2 \, r^4 \cdot 0.437~{\rm GeV}^4 \ \ \ \ \ {\rm for} \ N=3,
\nonumber \\
(\hat{y} F_\Theta M_\Theta)^2 &= \frac{N}{3} \, y_{r2}^2 \, r^4 \cdot 0.403~{\rm GeV}^4 \ \ \ \ \ {\rm for} \ N=4,
\nonumber \\
(\hat{y} F_\Theta M_\Theta)^2 &= \frac{N}{3} \, y_{r2}^2 \, r^4 \cdot 0.381~{\rm GeV}^4 \ \ \ \ \ {\rm for} \ N=5,
\nonumber \\
(\hat{y} F_\Theta M_\Theta)^2 &= \frac{N}{3} \, y_{r2}^2 \, r^4 \cdot 0.367~{\rm GeV}^4 \ \ \ \ \ {\rm for} \ N=6.
\label{ftheta}
\end{align}
The uncertainty of $(\hat{y} F_\Theta M_\Theta)^2$ due to the truncation of perturbative series is estimated to be $0.7^4\simeq25$\%,
 as the perturbative series in Eqs.~(\ref{3r3}),~(\ref{3r5}),~(\ref{6r3}),~(\ref{6r5}) converge at a rate about 0.7$^k$.
\\

\subsubsection{Ratio of the dynamical scale of the $SU(N)_T$ gauge theory and that of QCD}

From $M_\Theta$ and $\hat{y}F_\Theta$ evaluated in Eqs.~(\ref{mtheta}),~(\ref{ftheta}) and the matching condition~Eq.~(\ref{matching1}),
 we arrive at the following values of $r$:
\begin{align}
r &= \sqrt{\frac{3}{N}} \, \frac{1}{y_{r2}} \, 191 \ \ \ \ \ {\rm for \ }N=3,
\ \ \ \ \ \ \ 
r = \sqrt{\frac{3}{N}} \, \frac{1}{y_{r2}} \, 199 \ \ \ \ \ {\rm for \ }N=4,
\nonumber \\
r &= \sqrt{\frac{3}{N}} \, \frac{1}{y_{r2}} \, 204 \ \ \ \ \ {\rm for \ }N=5,
\ \ \ \ \ \ \ 
r = \sqrt{\frac{3}{N}} \, \frac{1}{y_{r2}} \, 208 \ \ \ \ \ {\rm for \ }N=6,
\label{ratio}
\end{align}
 with $y_{r2}$ denoting the $\bar{\chi}\psi H$ coupling constant $y$ at the scale $\mu=r\cdot2$~GeV in the $\overline{MS}$ scheme.
Here, the experimental value~\cite{pdg} $m_h=125.09$~GeV is used.
As a reference, we have $\Lambda_T \simeq $4.4~TeV for $y_{r2}=10^{-2}$ and $N=3$, with $\Lambda_{QCD}=0.23$~GeV.

Given the relation Eq.~(\ref{ratio}), the only model parameters are $y_{r2}$ and $N$.
Some may wonder that because $y_{r2}$ corresponds to the value of $y$ at the scale $r\cdot2$~GeV, which by itself contains $r$, 
 Eq.~(\ref{ratio}) might contain self-inconsistency.
As a matter of fact, the value of $y$ at one particular scale is involved in Eq.~(\ref{ratio}),
 and only after $r$ is determined from Eq.~(\ref{ratio}), can one calculate $y$ at different scales 
 using the RG equation with the initial condition $y(\mu=r\cdot2~{\rm GeV})=y_{r2}$.
\\

\section{Phenomenology of pseudo-Nambu-Goldstone bosons}

We are concerned with phenomenology of the pNG bosons that arise due to chiral symmetry breaking in the $SU(N)_T$ gauge theory.
They are associated with the spontaneous breaking of the axial $SU(3)_A\times U(1)_A$ symmetry,
 under which $\chi$ and $\psi$ transform as
\begin{align}
\left(
\begin{array}{c}
\chi  \\
\psi
\end{array}
\right)
 &\to \exp\left( i\theta_A^j \frac{\lambda^j}{2} \gamma_5 + i\theta_A \gamma_5 \right) \left(
\begin{array}{c}
\chi  \\
\psi
\end{array}
\right),
\end{align}
 where $\lambda^i$ $(i=1,2,...,8)$ are the Gell-Mann matrices, and $\theta_A^i$ $(i=1,2,...,8)$ and $\theta_A$ represent parameters of $SU(3)_A$ and $U(1)_A$ transformations.
Since the $U(1)_A$ symmetry is anomalous in the $SU(N)_T$ gauge theory, the corresponding pNG boson gains mass from instantons at the scale $4\pi \Lambda_T$, \textit{i.e.}, the dynamical scale multiplied by $4\pi$.
On the other hand, the $SU(3)_A$ symmetry is explicitly broken by the SM weak and hypercharge gauge interactions and the $\bar{\chi}\psi H$ Yukawa interaction.
Thus, the corresponding 8~pNG bosons are rendered massive by radiative corrections involving electroweak gauge bosons and elementary scalar $H$ as well as by the VEV of $H$,
 whose masses are at the scale $g_W \Lambda_T$ or $y \Lambda_T$.
Note that these masses are generated by the same mechanism as the charged pion-neutral pion mass difference, which stems from the electromagnetic interaction of quarks.
We eventually find that order $4\pi/g_W \sim 20$ mass hierarchy exists between the pNG boson of the $U(1)_A$ symmetry and those of the $SU(3)_A$ symmetry.
In the ensuing study of phenomenology, therefore, we ignore the former pNG boson and its mixing with the latter.
\\

\subsection{Mass spectrum of pseudo-Nambu-Goldstone bosons}

The mass matrix for the 8~pNG bosons of the $SU(3)_A$ symmetry is calculated with the Dashen's formula~\cite{dashen}
 in the leading order of the electroweak gauge couplings and the $\bar{\chi}\psi H$ Yukawa coupling as
\begin{align}
M^2_{ij} &= \frac{1}{f_\Pi^2} \langle 0 \vert [ Q_A^i, [Q_A^j, {\cal H}_{{\rm break}}]] \vert 0 \rangle \ \ \ (i,j=1,2,...,8),
\label{massmatrix}
\end{align}
 where $i,j$ label Gell-Mann matrices to which the pNG bosons correspond,
 $Q_A^i$ denotes the charge for a $SU(3)_A$ symmetry current $(\bar{\chi},\bar{\psi}) \gamma^\mu \gamma_5 \dfrac{\lambda^i}{2} \left(
\begin{array}{c}
 \chi \\
 \psi
\end{array}
\right)$, and $f_\Pi$ denotes the NG boson decay constant, which is approximated to be common for the 8~pNG bosons.
${\cal H}_{{\rm break}}$ is the effective Hamiltonian density that explicitly breaks the $SU(3)_A$ symmetry, which comprises two parts;
 one is obtained from the Lagrangian~Eq.(\ref{lagrangian1}) by contracting $W$, $Z$, photon and scalar fields
 with free field propagators; the other comes from the electroweak symmetry breaking VEV of $H$, $c_H v$.
In the calculation of ${\cal H}_{{\rm break}}$, we impose the unitary gauge for $W$ and $Z$ fields.
Also, we make the approximation that $c_H =1,s_H=0$.
Then only the physical Higgs field $h$ and $W,Z$ and photon fields contribute to ${\cal H}_{{\rm break}}$, 
 and it is expressed as
\begin{align}
&{\cal H}_{{\rm break}} = y \, \frac{v}{\sqrt{2}} \, (\bar{\chi}_2\psi + \bar{\psi}\chi_2)
\nonumber \\
&-\frac{i}{2} \, \frac{g_W^2}{2} \int{\rm d}^4x \, D^W_{\mu\nu}(x) \, \{ \bar{\chi}_1(x)\gamma^{\mu}\chi_2(x) \ \bar{\chi}_2(0)\gamma^{\nu}\chi_1(0) + \bar{\chi}_2(x)\gamma^{\mu}\chi_1(x) \ \bar{\chi}_1(0)\gamma^{\nu}\chi_2(0) \}
\nonumber \\
&- \frac{i}{2} \frac{g_Z^2}{4} \int{\rm d}^4x D^Z_{\mu\nu}(x)
\{ (c_W^2-bs_W^2)\bar{\chi}_1(x)\gamma^{\mu}\chi_1(x) - (c_W^2+bs_W^2)\bar{\chi}_2(x)\gamma^{\mu}\chi_2(x) + (1-b)s_W^2\bar{\psi}(x)\gamma^{\mu}\psi(x)\} 
\nonumber \\
& \ \ \ \ \ \ \ \ \ \ \ \ \ \ \ \ \times \{ (c_W^2-bs_W^2) \, \bar{\chi}_1(0)\gamma^{\nu}\chi_1(0) - (c_W^2+bs_W^2)\, \bar{\chi}_2(0)\gamma^{\nu}\chi_2(0) + (1-b)s_W^2 \, \bar{\psi}(0)\gamma^{\nu}\psi(0)\}
\nonumber \\
&- \frac{i}{2} \, \frac{e^2}{4} \int{\rm d}^4x \, D^\gamma_{\mu\nu}(x) \,
\{ (1+b) \, \bar{\chi}_1(x)\gamma^{\mu}\chi_1(x) - (1-b) \, \bar{\chi}_2(x)\gamma^{\mu}\chi_2(x) - (1-b) \, \bar{\psi}(x)\gamma^{\mu}\psi(x)\}
\nonumber \\
& \ \ \ \ \ \ \ \ \ \ \ \ \ \ \ \ \times \{ (1+b) \, \bar{\chi}_1(0)\gamma^{\nu}\chi_1(0) - (1-b) \, \bar{\chi}_2(0)\gamma^{\nu}\chi_2(0) - (1-b) \, \bar{\psi}(0)\gamma^{\nu}\psi(0)\} 
\nonumber \\
&- \frac{i}{2} \, \frac{y^2}{2} \int{\rm d}^4x \, D^h(x) \, \{ \bar{\chi}_2(x)\psi(x) + \bar{\psi}(x)\chi_2(x) \}\{ \bar{\chi}_2(0)\psi(0) + \bar{\psi}(0)\chi_2(0) \}
\label{effh}
\end{align}
 where $D^{W}_{\mu\nu}(x)$, $D^{Z}_{\mu\nu}(x)$, $D^\gamma_{\mu\nu}(x)$ and $D^h(x)$ denote the free field propagators for $W$, $Z$, photon and the physical Higgs field $h$, respectively,
 and in particular, the unitary gauge is chosen for $D^{W}_{\mu\nu}(x)$ and $D^{Z}_{\mu\nu}(x)$.
$c_W,s_W$ respectively denote the cosine and sine of the Weinberg angle.
By substituting ${\cal H}_{{\rm break}}$ into the Dashen's formula Eq.~(\ref{massmatrix}), 
 the non-vanishing components of the mass matrix are computed as
\begin{align}
M^2_{11} = M^2_{22} &= \frac{1}{f_\Pi^2} \left( \, g_W^2 \, {\cal C}^W + c_W^4g_Z^2 \, {\cal C}^Z + e^2 \, {\cal C}^\gamma + \frac{1}{4} \, {\cal F}_{12} \, \right),
\nonumber \\
M^2_{33} &= \frac{1}{f_\Pi^2} \left( \, 2g_W^2 \, {\cal C}^W + \frac{1}{4} \, {\cal F}_{38} \, \right),
\nonumber \\
M^2_{44} = M^2_{55} &= \frac{1}{f_\Pi^2} \left( \, \frac{1}{2}g_W^2 \, {\cal C}^W + \frac{(c_W^2-s_W^2)^2}{4}g_Z^2 \, {\cal C}^Z + e^2 \, {\cal C}^\gamma + \frac{1}{4} \, {\cal F}_{45} \, \right),
\nonumber \\
M^2_{66} = M^2_{77} &= \frac{1}{f_\Pi^2} \left( \, \frac{1}{2}g_W^2 \, {\cal C}^W + \frac{1}{4}g_Z^2 \, {\cal C}^Z + {\cal F}_{67} \, \right),
\nonumber \\
M^2_{88} &= \frac{1}{f_\Pi^2} \, \frac{1}{12} \, {\cal F}_{38},
\nonumber \\
M^2_{38} &= \frac{1}{f_\Pi^2} \, \frac{1}{4\sqrt{3}} \, {\cal F}_{38},
\nonumber \\
M^2_{14} = M^2_{25} = -M^2_{36} = \sqrt{3}M^2_{68} &= -\frac{1}{f_\Pi^2} \, \frac{v}{\sqrt{2}} \, \langle 0 \vert y \, \bar{f}f \vert 0 \rangle.
\label{pngmass}
\end{align}
Here, ${\cal C}^W,{\cal C}^Z,{\cal C}^\gamma$ are quantities defined with one massless Dirac fermion $f$ in the fundamental representation of the $SU(N)_T$ gauge group as
\begin{align}
{\cal C}^k &\equiv \frac{i}{2}\int{\rm d}^4x \, D^k_{\mu\nu}(x) \, \left[ \, \langle 0 \vert T \{ \bar{f}(x) \gamma^\mu f(x) \, \bar{f}(0) \gamma^\nu f(0) \} \vert 0 \rangle
- \langle 0 \vert T \{ \bar{f}(x) \gamma^\mu \gamma_5 f(x) \, \bar{f}(0) \gamma^\nu \gamma_5 f(0) \} \vert 0 \rangle \, \right]
\nonumber \\
&(k=W,Z,\gamma).
\label{cdef}
\end{align}
${\cal F}_{12}$, ${\cal F}_{38}$, ${\cal F}_{45}$, ${\cal F}_{67}$ are defined as
\begin{align}
{\cal F}_{12} &\equiv i\int{\rm d}^4x \, D^h(x) \, \left[ \, \langle 0 \vert T \{ y \, \bar{\chi}_2(x)\psi(x) \ y \, \bar{\psi}(0) \chi_2(0) \} \vert 0 \rangle - \langle 0 \vert T \{ y \, \bar{\chi}_1(x) i\gamma_5 \psi(x) \ y \, \bar{\psi}(0) i\gamma_5 \chi_1(0) \} \vert 0 \rangle \, \right],
\label{ddef12}
\\
{\cal F}_{38} &\equiv i\int{\rm d}^4x \, D^h(x) \, \left[ \, \langle 0 \vert T \{ y \, \bar{\chi}_2(x)\psi(x) \ y \, \bar{\psi}(0) \chi_2(0) \} \vert 0 \rangle - \langle 0 \vert T \{ y \, \bar{\chi}_2(x) i\gamma_5 \psi(x) \ y \, \bar{\psi}(0) i\gamma_5 \chi_2(0) \} \vert 0 \rangle \, \right],
\label{ddef3}
\\
{\cal F}_{45} &\equiv i\int{\rm d}^4x \, D^h(x) \, \left[ \, \langle 0 \vert T \{ y \, \bar{\chi}_2(x)\psi(x) \ y \, \bar{\psi}(0) \chi_2(0) \} \vert 0 \rangle - \langle 0 \vert T \{ y \, \bar{\chi}_2(x) i\gamma_5 \chi_1(x) \ y \, \bar{\chi}_1(0) i\gamma_5 \chi_2(0) \} \vert 0 \rangle \, \right],
\label{ddef45}
\\
{\cal F}_{67} &\equiv i\int{\rm d}^4x \, D^h(x) \, \left[ \, \langle 0 \vert T \{ y \, \bar{\chi}_2(x) \psi(x) \ y \, \bar{\psi}(0) \chi_2(0) \} \vert 0 \rangle \right.
\nonumber \\
&- \left. \frac{1}{2}\langle 0 \vert T \{ y \, \bar{\chi}_2(x) i\gamma_5 \chi_2(x) \ y \, \bar{\chi}_2(0) i\gamma_5 \chi_2(0) \} \vert 0 \rangle
- \frac{1}{2}\langle 0 \vert T \{ y \, \bar{\psi}(x) i\gamma_5 \psi(x) \ y \, \bar{\psi}(0) i\gamma_5 \psi(0) \} \vert 0 \rangle \, \right].
\label{ddef67}
\end{align}
Finally, in the last line of Eq.~(\ref{pngmass}), $\langle 0 \vert y \, \bar{f}f \vert 0 \rangle$ collectively denotes the fermion bilinear condensate with the coupling constant $y$, $\langle 0 \vert y \, \bar{\chi}_1\chi_1 \vert 0 \rangle=\langle 0 \vert y \, \bar{\chi}_2\chi_2 \vert 0 \rangle=\langle 0 \vert y \, \bar{\psi}\psi \vert 0 \rangle$.

In the rest of the subsection, we evaluate the pNG boson masses~Eq.~(\ref{pngmass}) by analogy with QCD.
This is done by equating ${\cal C}^W,{\cal C}^Z,{\cal C}^\gamma,{\cal F}_1,{\cal F}_2$ and the pNG boson decay constant $f_\Pi$ with certain quantities in QCD, rescaled by the dynamical scale ratio $r$ as well as $N/N_c$.
Then $r$ is rendered into the coupling constant $y_{r2}$ and $N$ through Eq.~(\ref{ratio}), by which the pNG boson masses are expressed solely in terms of $y_{r2}$ and $N$.
\\

\subsubsection{Evaluation of $f_\Pi$}

$f_\Pi$ is evaluated from the pion decay constant in the chiral-limit QCD, $f_\pi^{{\rm chiral}}$, obtained in Ref.~\cite{durr}
 by fitting a lattice simulation with the next-to-leading order chiral perturbation theory~\cite{leutwyler}.
It is given by
\begin{align}
f_\pi^{{\rm chiral}} &= 0.08678~{\rm GeV}.
\label{lattice}
\end{align}
Reminding that the NG boson decay constant scales by $\sqrt{N/N_c}$ as well as $r$,
 we obtain the following estimate:
\begin{align}
f_\Pi &= \sqrt{\frac{N}{N_c}} \, r \, f_\pi^{{\rm chiral}} = \sqrt{\frac{N}{N_c}} \, r \cdot 0.08676~{\rm GeV}.
\end{align}
\\

\subsubsection{Evaluation of ${\cal C}^W,{\cal C}^Z,{\cal C}^\gamma$}

${\cal C}^W,{\cal C}^Z,{\cal C}^\gamma$ in Eq.~(\ref{cdef}) can be estimated from the mass difference between the charged and neutral pions in QCD.
Recall that in QCD, this mass difference stems mainly from the electromagnetic interaction.
Substituting the effective Hamiltonian density for the electromagnetic interaction of quarks into the Dashen's formula, we compute 
 the pion mass difference $m_{\pi^\pm}^2-m_{\pi^0}^2$ to be
\begin{align}
&m_{\pi^\pm}^2 - m_{\pi^0}^2 
\nonumber \\
&= \frac{e^2}{f_\pi^2} \frac{i}{2}\int{\rm d}^4x \, D_{\mu\nu}^\gamma(x) \left[ \, \langle 0 \vert T \{ \bar{q}(x) \gamma^\mu q(x) \, \bar{q}(0) \gamma^\nu q(0) \} \vert 0 \rangle
- \langle 0 \vert T \{ \bar{q}(x) \gamma^\mu \gamma_5 q(x) \, \bar{q}(0) \gamma^\nu \gamma_5 q(0) \} \vert 0 \rangle \, \right],
\label{pionmassdifference}
\end{align}
 where $q$ denotes up and down quarks, and $f_\pi$ denotes the pion decay constant in real QCD.
Analogy between QCD and the $SU(N)_T$ gauge theory gives that
 the integral on the right hand side of Eq.~(\ref{pionmassdifference}) is $(N_c/N){\cal C}^\gamma/r^4$,
 where the factor $N/N_c$ enters because the correlation function scales by $N/N_c$.
We thus find the following relation between ${\cal C}^\gamma$ and the pion mass difference:
\begin{align}
{\cal C}^\gamma &= r^4 \, \frac{N}{N_c} \frac{1}{e^2} f_\pi^2(m_{\pi^\pm}^2 - m_{\pi^0}^2).
\label{cestimate}
\end{align}
Using experimental central values~\cite{pdg} $e^2=4\pi/137.0$, $m_{\pi^\pm}=0.13957018$~GeV, $m_{\pi^0}=0.1349766$~GeV and $f_\pi=0.921$~GeV,
 we obtain
\begin{align}
{\cal C}^\gamma &= \frac{N}{N_c} \, r^4 \, (0.104~{\rm GeV})^4. 
\label{cnumestimate}
\end{align}

Regarding ${\cal C}^W$ and ${\cal C}^Z$, notice that they are independent of the gauge choice for $W$ and $Z$ fields, because
 the correlation function for the axial current $\bar{f} \gamma^\mu \gamma_5 f$, as with the vector current,
 is proportional to $g_{\mu\nu}p^2 - p_\mu p_\nu$ ($p^\mu$ denotes the momentum of the correlation function) 
 in the zeroth order of the electroweak gauge couplings and $\bar{\chi}\psi H$ Yukawa coupling due to the absence of $\chi,\psi$ current mass.
Also, in the limit with $y_{r2} \ll 1$, the dynamical scale of the $SU(N)_T$ gauge theory is much larger than the electroweak scale
 and hence the $W$ and $Z$ boson masses can be ignored in the calculation of ${\cal C}^W$ and ${\cal C}^Z$.
Therefore, ${\cal C}^W$, ${\cal C}^Z$ and ${\cal C}^\gamma$ can be calculated with the identical free field propagator for the gauge field and we find
\begin{align}
{\cal C}^W &= {\cal C}^Z = {\cal C}^\gamma.
\end{align}
\\

\subsubsection{Evaluation of ${\cal F}_{12},{\cal F}_{38},{\cal F}_{45},{\cal F}_{67}$}

Since ${\cal F}_{12}$, ${\cal F}_{38}$, ${\cal F}_{45}$ and ${\cal F}_{67}$ have no corresponding quantities in QCD, we calculate them explicitly.
We commence the calculation from ${\cal F}_{38}$.
It can be recast in the following form:
\begin{align}
{\cal F}_{38} &= \frac{i}{(2\pi)^4}\int{\rm d}^4p \, \frac{1}{p^2-m_h^2}\left\{ \, \Pi_{y\bar{\chi}_2\psi}(p^2) - \Pi_{y\bar{\chi}_2i\gamma_5\psi}(p^2) \, \right\}
\nonumber \\
&= \frac{1}{16\pi^2}\int_0^\infty{\rm d}(p_E^2) \, \frac{p_E^2}{p_E^2+m_h^2}\left\{ \, \Pi_{y\bar{\chi}_2\psi}(-p_E^2) - \Pi_{y\bar{\chi}_2i\gamma_5\psi}(-p_E^2) \, \right\},
\label{f3calc}
\end{align}
 where $p_E$ is the Euclidean momentum with $p^0=ip_E^0, p^{1,2,3}=p_E^{1,2,3}$, and the scalar and pseudoscalar current correlation functions $\Pi_{y\bar{\chi}_2\psi}$ and $\Pi_{y\bar{\chi}_2i\gamma_5\psi}$ are given by
\begin{align}
\Pi_{y\bar{\chi}_2\psi}(p^2) &= i\int{\rm d}^4x \, e^{ipx} \langle 0 \vert T\left\{ y \, \bar{\chi}_2(x)\psi(x) \ y \, \bar{\psi}(0)\chi_2(0) \right\} \vert 0 \rangle,
\nonumber \\
\Pi_{y\bar{\chi}_2i\gamma_5\psi}(p^2) &= i\int{\rm d}^4x \, e^{ipx} \langle 0 \vert T\left\{ y \, \bar{\chi}_2(x)i\gamma_5\psi(x) \ y \, \bar{\psi}(0)i\gamma_5\chi_2(0) \right\} \vert 0 \rangle.
\end{align}
${\cal F}_{38}$ is thus calculated from the difference between the scalar and pseudoscalar correlation functions for space-like momenta $p^2\leq0$.
The correlation functions for space-like momenta are connected to their imaginary parts for time-like momenta by the dispersion relation.
As the operator product expansion~\cite{qcdsumrules} yields $\Pi_{y\bar{\chi}_2\psi}(-p_E^2) - \Pi_{y\bar{\chi}_2i\gamma_5\psi}(-p_E^2) = O(1/p_E^4)$ for $p_E^2 \to \infty$
 (recall that fermion bilinear condensate does not enter into the correlation function due to the absence of $\chi,\psi$ current mass),
 no subtraction term is needed for the correlation function difference and we find
\begin{align}
\Pi_{y\bar{\chi}_2\psi}(-p_E^2) - \Pi_{y\bar{\chi}_2i\gamma_5\psi}(-p_E^2) &= \frac{1}{\pi}\int_0^\infty{\rm d}s \, \frac{{\rm Im}\Pi_{y\bar{\chi}_2\psi}(s) - {\rm Im}\Pi_{y\bar{\chi}_2i\gamma_5\psi}(s)}{s+p_E^2}.
\label{dispersion2}
\end{align}

The imaginary part of the correlation function difference, ${\rm Im}\Pi_{y\bar{\chi}_2\psi}(s)-{\rm Im}\Pi_{y\bar{\chi}_2i\gamma_5\psi}(s)$,
 corresponds to the difference between the spectral densities of scalar and pseudoscalar bound states.
We infer the form of the spectral density difference from the fact that chiral symmetry is restored for large momenta $s \to \infty$.
This implies that the scalar and pseudoscalar spectral densities coincide for heavy bound states
 and their difference is described with several light bound states.
We thus assume that the spectral density difference is approximated by contributions from the scalar meson $\Theta$,
 the pNG boson $\frac{1}{\sqrt{2}}(\Pi^6 \pm i\Pi^7)$ and one heavier pseudo-scalar meson, which we denote by $\Pi'$.
\footnote{
$\Pi'$ may correspond to the unestablished $K(1460)$ meson.
}
It is further assumed that $\Theta$ and $\Pi'$ meson resonances are described by the relativistic Breit-Wigner function,
 whereas the pNG boson contribution is represented by a delta function at $s=0$ because it is massless in the zeroth order of the electroweak gauge couplings and the Yukawa coupling $y$.
Given these approximations, we arrive at the following spectral density difference:
\begin{align}
&\frac{1}{\pi}\left\{ {\rm Im}\Pi_{y\bar{\chi}_2\psi}(s)-{\rm Im}\Pi_{y\bar{\chi}_2i\gamma_5\psi}(s) \right\} 
\nonumber \\
&=
 (\hat{y} F_\Theta M_\Theta)^2 \, \frac{1}{\pi}\frac{M_\Theta \Gamma_\Theta}{(s-M_\Theta^2)^2+M_\Theta^2 \Gamma_\Theta^2} 
 - \hat{y}^2 G_{\Pi}^2 \, \delta(s) - \hat{y}^2 G_{\Pi'}^2 \, \frac{1}{\pi}\frac{M_{\Pi'} \Gamma_{\Pi'}}{(s-M_{\Pi'}^2)^2+M_{\Pi'}^2 \Gamma_{\Pi'}^2} 
\label{spsspectral}
\end{align}
 where $M_{\Pi'}$ and $\Gamma_{\Pi'}$ respectively denote the mass and width of $\Pi'$ meson, and $\hat{y}G_{\Pi}$ and $\hat{y}G_{\Pi'}$ are defined as
\begin{align}
\langle 0 \vert y \, \bar{\chi}_2(x) i\gamma_5 \psi(x) \vert \, \frac{1}{\sqrt{2}}(\Pi^6+i\Pi^7)(p) \, \rangle &\equiv \hat{y} G_{\Pi} \, e^{-ipx},
\label{gpipre} \\
\langle 0 \vert y \, \bar{\chi}_2(x) i\gamma_5 \psi(x) \vert \Pi'(p) \rangle &\equiv \hat{y} G_{\Pi'} \, e^{-ipx},
\end{align}
 with $\hat{y}$ denoting the renormalization group invariant quantity made from the coupling constant $y$.

Substituting the spectral density difference Eq.~(\ref{spsspectral}) into the dispersion relation Eq.~(\ref{dispersion2}),
 we compute the correlation function difference to be
\begin{align}
&\Pi_{y\bar{\chi}_2\psi}(-p_E^2) - \Pi_{y\bar{\chi}_2i\gamma_5\psi}(-p_E^2)
\nonumber \\
&= 
(\hat{y} F_\Theta M_\Theta)^2 \, \int_0^\infty {\rm d}s \, \frac{1}{s+p_E^2}\frac{1}{\pi}\frac{M_\Theta \Gamma_\Theta}{(s-M_\Theta^2)^2+M_\Theta^2 \Gamma_\Theta^2} - \hat{y}^2 G_{\Pi}^2 \frac{1}{p_E^2}
\nonumber \\
&- \hat{y}^2 G_{\Pi'}^2 \int_0^\infty {\rm d}s \, \frac{1}{s+p_E^2}\frac{1}{\pi}\frac{M_{\Pi'} \Gamma_{\Pi'}}{(s-M_{\Pi'}^2)^2+M_{\Pi'}^2 \Gamma_{\Pi'}^2}.
\label{spscorrelator}
\end{align}
${\cal F}_{38}$ is calculated from the above correlation function difference through Eq.~(\ref{f3calc}).
Before that, we remind that in the limit with $p_E^2\to \infty$,
 the correlation function difference Eq.~(\ref{spscorrelator}) asymptotes to the VEV of the operator product expansion calculated in the $SU(N)_T$ gauge theory, as the expansion is reliable in this limit.
The calculation in Ref.~\cite{qcdsumrules} tells us that $O(1)$ and $O(1/p_E^2)$ terms in $\Pi_{y\bar{\chi}_1\psi}(-p_E^2) - \Pi_{y\bar{\chi}_1i\gamma_5\psi}(-p_E^2)$
 vanish for $p_E^2\to \infty$ (note that fermion bilinear condensate does not appear in our case), yielding the following constraint:
\begin{align}
&\frac{1}{\pi}\int_0^\infty{\rm d}s \, \left\{ \, (\hat{y} F_\Theta M_\Theta)^2 \frac{M_\Theta \Gamma_\Theta}{(s-M_\Theta^2)^2+M_\Theta^2 \Gamma_\Theta^2} - \hat{y}^2 G_{\Pi'}^2 \frac{M_{\Pi'} \Gamma_{\Pi'}}{(s-M_{\Pi'}^2)^2+M_{\Pi'}^2 \Gamma_{\Pi'}^2} \, \right\}  - \hat{y}^2 G_{\Pi}^2 = 0.
\label{gpipconstraint}
\end{align}
By virtue of the above constraint, the integral over $p_E^2$ in Eq.~(\ref{f3calc}) is convergent and we obtain
\begin{align}
{\cal F}_{38} &= -\frac{1}{16\pi^2}\int_0^\infty{\rm d}s \, \frac{s}{s-m_h^2}\log\left(\frac{s}{m_h^2}\right) \frac{1}{\pi}
\nonumber \\
&\times \left\{ \,
(\hat{y} F_\Theta M_\Theta)^2 \frac{M_\Theta \Gamma_\Theta}{(s-M_\Theta^2)^2+M_\Theta^2 \Gamma_\Theta^2} - \hat{y}^2 G_{\Pi'}^2 \frac{M_{\Pi'} \Gamma_{\Pi'}}{(s-M_{\Pi'}^2)^2+M_{\Pi'}^2 \Gamma_{\Pi'}^2} \, \right\}.
\label{f1}
\end{align}
We evaluate ${\cal F}_{38}$ by making the following two assumptions:
First, we assume that $M_{\Pi'}$ is located at the same scale as $M_\Theta$, namely, $M_{\Pi'}\sim M_\Theta$ is assumed.
Second, the ratio of the total width over the mass for $\Pi'$ is similar to that for $\Theta$,
 \textit{i.e.}, the $\Pi'$ resonance is as narrow as the $\Theta$ resonance.
Then, since $\Gamma_{\Pi'}/M_{\Pi'} \simeq \Gamma_\Theta/M_\Theta \simeq 0.2$, the integrand has a significant peak in the region with $s \sim M_\Theta \sim M_{\Pi'}$
 and the integral in Eq.~(\ref{f1}) is dominated by contributions from this region.
It follows that in the limit with $y_{r2} \ll 1$, we have $M_{\Pi'} \sim M_\Theta \gg m_h$ and $\vert \log(M_\Theta^2/M_{\Pi'}^2) \vert \ll \log(M_\Theta^2/m_h^2)$,
 and are thus allowed to make the approximations $s/(s-m_h^2) = 1$ and $\log(s/m_h^2) = \log(M_\Theta^2/m_h^2)$ in the integrand.
Under these approximations, ${\cal F}_{38}$ is evaluated as
\begin{align}
{\cal F}_{38} &= -\frac{1}{16\pi^2}\log\left(\frac{M_\Theta^2}{m_h^2}\right) \frac{1}{\pi} \int_0^\infty{\rm d}s \, \left\{ \,
(\hat{y} F_\Theta M_\Theta)^2 \frac{M_\Theta \Gamma_\Theta}{(s-M_\Theta^2)^2+M_\Theta^2 \Gamma_\Theta^2} - \hat{y}^2 G_{\Pi'}^2 \frac{M_{\Pi'} \Gamma_{\Pi'}}{(s-M_{\Pi'}^2)^2+M_{\Pi'}^2 \Gamma_{\Pi'}^2} \, \right\}
\label{f3app}
\end{align}
Again making use of the constraint Eq.~(\ref{gpipconstraint}), we arrive at
\begin{align}
{\cal F}_{38} &= -\frac{1}{16\pi^2}\log\left(\frac{M_\Theta^2}{m_h^2}\right) \, \hat{y}^2 G_{\Pi}^2.
\label{f3app2}
\end{align}

We will express the term $\hat{y}G_\Pi$ as a combination of certain quantities in QCD multiplied by the dynamical scale ratio $r$ and $N$.
As a first step, we prove that $\hat{y}G_{\Pi}$ is linked to the fermion bilinear condensate and the NG boson decay constant in the following manner: 
\begin{align}
\hat{y} G_{\Pi} &= \langle 0 \vert y \, \bar{\chi}_1(x) i\gamma_5 \psi(x) \vert \, \frac{1}{\sqrt{2}}(\Pi^6+i\Pi^7)(p) \, \rangle \, e^{ipx}
\nonumber \\
&= -\sqrt{2} \, \frac{\langle 0 \vert y \, \bar{f}f \vert 0 \rangle }{f_\Pi},
\label{gpi}
\end{align}
 with $f$ representing one massless Dirac fermion in the fundamental representation of the $SU(N)_T$ gauge group.
To prove Eq.~(\ref{gpi}), consider adding an infinitesimal current mass $y \delta v$ to the $\psi$ field as $-\delta{\cal L}=y\delta v \bar{\psi}\psi$, with $\delta v$ being an infinitesimal RG invariant constant.
Its contribution to the $\Pi^6,\Pi^7$ mass is computed as
\begin{align}
\delta M_{66}^2 = \delta M_{77}^2 &= -\frac{1}{f_\Pi^2} \langle 0 \vert y\delta v \, \bar{f}f \vert 0 \rangle.
\label{infinitesimalmass}
\end{align}
On the other hand, the NG boson low-energy theorem relates the axial vector current and the one-NG-boson state in the following way:
\begin{align}
\langle 0 \vert \frac{1}{\sqrt{2}}\bar{\chi}_2(x) \gamma_\mu \gamma_5 \psi(x) \vert \, \frac{1}{\sqrt{2}}(\Pi^6+i\Pi^7)(p) \, \rangle &= if_\Pi \, p_\mu \, e^{-ipx}.
\label{low}
\end{align}
Taking total derivative on both sides of Eq.~(\ref{low}), and utilizing the equation of motion for $\psi$ field operator with the electroweak and $\bar{\chi}\psi H$ Yukawa interactions ignored, which reads $(i\partial_\mu \gamma^\mu - y \, \delta v)\psi=0$, we obtain
\begin{align}
\langle 0 \vert \frac{y \, \delta v}{\sqrt{2}}\bar{\chi}_2(x) i\gamma_5 \psi(x) \vert \, \frac{1}{\sqrt{2}}(\Pi^6+i\Pi^7)(p) \, \rangle &= f_\Pi \, p^2 \, e^{-ipx} = f_\Pi \, \delta M_{66}^2 \, e^{-ipx}.
\label{low2}
\end{align}
Substituting the mass formula Eq.~(\ref{infinitesimalmass}) into the right-hand side of Eq.~(\ref{low2}) and then 
 taking a derivative with $\delta v$ on both sides, we find
\begin{align}
\langle 0 \vert \frac{y}{\sqrt{2}}\bar{\chi}_2(x) i\gamma_5 \psi(x) \vert \, \frac{1}{\sqrt{2}}(\Pi^6+i\Pi^7)(p) \, \rangle &= -\frac{1}{f_\Pi} \langle 0 \vert y \, \bar{f}f \vert 0 \rangle \, e^{-ipx},
\label{low3}
\end{align}
 reproducing Eq.~(\ref{gpi}) as desired.
As a second step, we compare the fermion bilinear condensate $\langle 0 \vert y \, \bar{f}f \vert 0 \rangle$ and the NG boson decay constant $f_\Pi$ to corresponding quantities in QCD,
 for which we employ the pion decay constant and the quark bilinear condensate in the chiral-limit QCD obtained in Ref.~\cite{durr}
 by fitting a lattice simulation with the next-to-leading order chiral perturbation theory~\cite{leutwyler},
 which read
\begin{align}
f_\pi^{{\rm chiral}} &= 0.08678~{\rm GeV},
\label{lattice1}
\\
-\frac{ \langle 0 \vert (m_u+m_d) \bar{q}q \vert 0 \rangle^{{\rm chiral}} }{ (f_\pi^{{\rm chiral}})^2 } &= 0.01861~{\rm GeV}^2.
\label{lattice2}
\end{align}
Here, the quark bilinear condensate in Eq.~(\ref{lattice2}) is defined in terms of the product of a quark bilinear operator and the current quark mass, which is the sum of up and down quark current mass $(m_u+m_d)$.
This product is independent of the wavefunction renormalization in QCD.
Eq.~(\ref{lattice2}) is computed by first tuning $(m_u+m_d)$ to reproduce the real pion mass-pion decay constant ratio $m_{\pi^0}/f_{\pi}$.
For this value of $(m_u+m_d)$, the lattice spacing is determined from the experimental value of pion mass.
Then the quark bilinear condensate is extrapolated for $(m_u+m_d)=0$.
To translate $\langle 0 \vert (m_u+m_d) \bar{q}q \vert 0 \rangle^{{\rm chiral}}$ into $\langle 0 \vert y \, \bar{f}f \vert 0 \rangle$,
 we exploit the fact that the current quark mass $(m_u+m_d)$ in QCD and the coupling constant $y$ in the $SU(N)_T$ gauge theory are renormalized in the same way with respect to the QCD and $SU(N)_T$ gauge coupling constants.
For $N=3$, the correspondence is exact and we have $(m_u+m_d)(\mu')/(m_u+m_d)(\mu) = y(r \mu')/y(r \mu)$ for any two scales $\mu,\mu'$.
For $N\geq4$, we still have an approximate relation $(m_u+m_d)(\mu')/(m_u+m_d)(\mu) \simeq y(r \mu')/y(r \mu)$.
This is because the gauge coupling roughly scales as $\alpha_T(r\mu) \simeq (N_c/N)\alpha_s(\mu)$ whereas the Casimir operator $C_F$ scales as $C_F(N)\simeq (N/N_c)C_F(N_c)$ and $C_A$ scales as $C_A(N)=(N/N_c)C_A(N_c)$.
The $\chi,\psi$ loop contribution to the gauge field propagator does not scale in this way, but its impact is subdominant compared to contributions from the gauge field and ghost field loops.
Therefore, the factor $N/N_c$ mostly cancels and the radiative corrections to $(m_u+m_d)$ and $y$ are similar in numerical values.
Accordingly, $\langle 0 \vert y \, \bar{f}f \vert 0 \rangle$ is estimated from $\langle 0 \vert (m_u+m_d) \bar{q}q \vert 0 \rangle^{{\rm chiral}}$ as
\begin{align}
\langle 0 \vert y \, \bar{f}f \vert 0 \rangle &\simeq r^3 \, \frac{N}{N_c} \, y(r \mu) \frac{ \langle 0 \vert (m_u+m_d) \bar{q}q \vert 0 \rangle^{{\rm chiral}} }{(m_u+m_d)(\mu)},
\label{yrelation}
\end{align}
 where the factor $N/N_c$ is because the fermion bilinear condensate scales in the same manner as a correlation function for $SU(N)_T$-singlet operators.
We quote the numerical value of $(m_u+m_d)(\mu)$ for $\mu=2$~GeV in the $\overline{MS}$ scheme obtained by a lattice simulation in Ref.~\cite{durr2}, which reads
\begin{align}
\frac{1}{2}(m_u+m_d)(\mu=2~{\rm GeV}) &= 0.003469~{\rm GeV}.
\label{lattice3}
\end{align}
Assembling the lattice results Eqs.~(\ref{lattice1}),~(\ref{lattice2}),~(\ref{lattice3}), the relation Eq.~(\ref{yrelation}) and the value of $f_\Pi$ evaluated in Section~3.1.1,
 we numerically evaluate ${\cal F}_{38}$ Eq.~(\ref{f3app2}) as
\begin{align}
{\cal F}_{38} &= -\frac{1}{16\pi^2}\log\left(\frac{M_\Theta^2}{m_h^2}\right) \, \left(-\sqrt{2}\frac{\langle 0 \vert y \, \bar{f}f \vert 0 \rangle }{f_\Pi}\right)^2
\nonumber \\
&= -\frac{1}{16\pi^2}\log\left(\frac{r^2 \, m_{K_0^*(1430)}^2}{m_h^2}\right) \, \left(-\sqrt{2}y(r\mu)\frac{\langle 0 \vert (m_u+m_d)\bar{q}q \vert 0 \rangle^{{\rm chiral}} }{(m_u+m_d)(\mu)} \, \frac{1}{f_\pi^{{\rm chiral}}}\right)^2 \, \frac{N}{N_c}r^4
\nonumber \\
&= -\frac{N}{N_c} \, y_{r2}^2 \, r^4 \, \log(0.000130r^2) \, (0.162~{\rm GeV})^4,              
\end{align}
 where $y_{r2}$ is the coupling constant $y$ evaluated at $\mu=r\cdot2$~GeV scale in the $\overline{MS}$ scheme, which coincides with what has appeared in Section~2.2.

${\cal F}_{12}$ and ${\cal F}_{45}$ can be derived by the same reasoning as ${\cal F}_{38}$, expect that the pNG boson $(\Pi^6\pm i\Pi^7)$ should be exchanged with $(\Pi^1\pm i\Pi^2)$ or $(\Pi^4\pm i\Pi^5)$.
Since these pNG bosons can all be represented by a delta function at $s=0$, the calculations are identical and we obtain
\begin{align}
{\cal F}_{12} &={\cal F}_{45}={\cal F}_{38}.
\end{align}
On the other hand, the evaluation of ${\cal F}_{67}$ does not proceed in the same way as ${\cal F}_{38}$,
 because ${\cal F}_{67}$ involves the pNG boson associated with an anomalous axial current $\bar{\psi} i \gamma_\mu \gamma_5 \psi + \bar{\chi}_2 i \gamma_\mu \gamma_5 \chi_2$.
We speculate ${\cal F}_{67} \sim \frac{N}{N_c} \, y_{2r}^2 \, \Lambda_T^4$ and postpone its evaluation for future works.
\\

\subsubsection{Evaluation of the pseudo-Nambu-Goldstone boson mass}

Now that $f_\Pi,{\cal C}^\gamma,{\cal C}^W,{\cal C}^Z,{\cal F}_{38},{\cal F}_{12},{\cal F}_{45}$ are written with the dynamical scale ratio $r$ as well as $N$,
 we are in position to express the pNG boson mass matrix Eq.~(\ref{pngmass}) solely in terms of $y_{r2}$ and $N$, using Eq.~(\ref{ratio}).
We restrict ourselves to the limit with $y_{r2} \ll g_W$.
Also, we ignore the tiny numerical difference in the evaluation of $r$ for $N=3,4,5,6$
 and approximate it as
\begin{align}
r &= \sqrt{\frac{3}{N}} \, \frac{1}{y_{r2}} \, 2.0\times10^2 \ \ \ \ \ {\rm for \ all \ }N.
\end{align}

When $y_{r2} \ll g_W$, we have $g_W^2{\cal C}^W \gg {\cal F}_{38} = {\cal F}_{12} = {\cal F}_{45}$, $g_W^2{\cal C}^W \gg {\cal F}_{67}$ and $g_W^2{\cal C}^W \gg v \hat{y} \langle 0 \vert \bar{f}f \vert 0 \rangle$,
 which gives that off-diagonal mass terms $M^2_{ij} \, (1\leq i \neq j \leq 8)$ can be ignored in comparion to the diagonal ones
\footnote{
$M^2_{38}$ can be ignored despite $M^2_{88}$ being smaller than $M^2_{38}$, because $M^2_{33} \gg M^2_{38}$.
}
 and the terms ${\cal F}_{12}, \, {\cal F}_{38}, \, {\cal F}_{45}$ and ${\cal F}_{67}$ in $M^2_{11}, M^2_{22},...,M^2_{77}$ can further be discarded.
The pNG boson masses are thus found to be
\begin{align}
M^2_{11} = M^2_{22} = M^2_{33} = \frac{1}{f_\Pi^2} 2g_W^2 \, {\cal C}^\gamma &= \frac{3}{N}\frac{2g_W^2}{y^2_{r2}} \, (25~{\rm GeV})^2, 
\nonumber \\
M^2_{44} = M^2_{55} = M^2_{66} = M^2_{77} = \frac{1}{f_\Pi^2} \left(\frac{g_W^2}{2} + \frac{g_Z^2}{4} \right) {\cal C}^\gamma &= \frac{3}{N}\frac{1}{y^2_{r2}} \left(\frac{g_W^2}{2} + \frac{g_Z^2}{4} \right) \, (25~{\rm GeV})^2,
\nonumber \\
M^2_{88} = \frac{1}{f_\Pi^2} \, \frac{1}{12} \, {\cal F}_{38} &= -\frac{3}{N} \log\left(5.2 \, \frac{3}{N} \frac{1}{y^2_{r2}}\right) \, (17~{\rm GeV})^2. 
\label{pngmass2}
\end{align}
Note that ${\cal F}_{38}$ and hence $M^2_{88}$ are negative, in contrast with ${\cal C}^\gamma$ being positive.
A physical interpretation for this is that because scalar exchange force is always attractive, the energy of a system containing a scalar interaction ($\Pi^8$ meson in this case) diminishes, which manifests itself as a negative radiative correction to the mass.
On the other hand, vector exchange force between the same charge is repulsive and thus contributes positively to the energy, and hence to the mass, of a system containing this type of interaction.

Since $M^2_{88}$ is negative, the $\Pi^8$ field acquires a VEV.
The vacuum can be stabilized by a quartic coupling for the $\Pi^8$ field, which is generated radiatively from the $\bar{\chi}\psi H$ Yukawa interaction that explicitly violates the $\lambda^8$ component of the $SU(3)_A$ symmetry.
The quartic coupling is proportional to
\begin{align}
\int{\rm d}^4x  \, D^h(x) \, \langle 0 \vert \, y \bar{\chi}_2\psi(x) \, y \bar{\psi}\chi_2(0) \, \vert \Pi^8 \Pi^8 \Pi^8 \Pi^8 \rangle
\label{quartic}
\end{align}
 in the leading order of the coupling constant $y$.
The quartic coupling is na\"ively estimated as $(N_c/N)y_{r2}^2$, giving rise to the term
\begin{align}
-{\cal L} &\supset c \, \frac{N_c}{N}y_{r2}^2 \, (\Pi^8)^4,
\end{align}
 where $c$ is a $O(1)$ constant, and the factor $N_c/N$ originates from the fact that when the $\Pi^8$-meson creation operator asymptotes to a current operator, a factor $\sqrt{N_c/N}$ appears,
 and the correlation function for $SU(N)_T$-singlet current operators scales by $N/N_c$.
If positive, the quartic coupling is responsible for vacuum stabilization and the VEV of $\Pi^8$ satisfies
\begin{align}
\langle \Pi^8 \rangle &\sim \sqrt{ \frac{-M^2_{88}}{\frac{N_c}{N}y_{r2}^2} } = \frac{1}{y_{r2}} \sqrt{\log\left(5.2 \, \frac{3}{N} \frac{1}{y^2_{r2}}\right)} \, 17~{\rm GeV}.
\label{pi8vev}
\end{align}
Phenomenologically, the VEV breaks parity ($P$) and charge-conjugation-parity ($CP$) symmetries, and brings about
 a $CP$-violating mixing of the physical Higgs field $h$ and the pNG boson $\Pi^6$,
\begin{align}
-{\cal L} &\supset h \, \Pi^6 \, \langle \, \Pi^6 \, \vert ( \, y \bar{\chi}_2\psi(0)+ y \bar{\psi}\chi_2(0) \, ) \vert \, \Pi^8 \, \rangle \, \langle \Pi^8 \rangle.
\end{align}
Since $\Pi^6$ does not couple to SM fermions, the above mixing should be detected as a suppression on the coupling of the observed Higgs particle to SM fermions as compared to the SM.
To gain information on the mixing angle for $h$ and $\Pi^6$, we estimate the mixing term based on its mass dimension and $y$ dependence as
\begin{align}
-{\cal L} &\supset \sim \ h \, \Pi^6 \, y_{r2} \Lambda_T \, \langle \Pi^8 \rangle,
\end{align}
 which yields the following mass matrix for $h$ and $\Pi^6$:
\begin{align}
&-{\cal L}\supset \sim \ \frac{1}{2}\left(
\begin{array}{cc}
\Pi^6 & h
\end{array}
\right)
\left(
\begin{array}{cc}
M_{66}^2 & y_{r2} \Lambda_T \, \langle \Pi^8 \rangle \\
y_{r2} \Lambda_T \, \langle \Pi^8 \rangle & m_h^2
\end{array}
\right)
\left(
\begin{array}{c}
\Pi^6 \\
h 
\end{array}
\right)
\nonumber \\
&= \frac{1}{2}\left(
\begin{array}{cc}
\Pi^6 & h
\end{array}
\right)
\left(
\begin{array}{cc}
\frac{3}{N}\frac{1}{y^2_{r2}} \left(\frac{g_W^2}{2} + \frac{g_Z^2}{4} \right) \, (25~{\rm GeV})^2 & \frac{1}{y_{r2}} \sqrt{\frac{3}{N}\log\left(5.2 \, \frac{3}{N} \frac{1}{y^2_{r2}}\right)} \, 46\cdot17~{\rm GeV}^2
\nonumber \\
\frac{1}{y_{r2}} \sqrt{\frac{3}{N}\log\left(5.2 \, \frac{3}{N} \frac{1}{y^2_{r2}}\right)} \, 46\cdot17~{\rm GeV}^2 & m_h^2
\end{array}
\right)
\left(
\begin{array}{c}
\Pi^6 \\
h 
\end{array}
\right),
\end{align}
 where we have used the estimate for $\langle \Pi^8 \rangle$ in Eq.~(\ref{pi8vev}) and the relation $y_{r2} \Lambda_T = \sqrt{3/N} (2.0\times10^2) \Lambda_{QCD}$ as well as $\Lambda_{QCD}=0.23$~GeV.
We find that in the limit with $y_{r2} \ll g_W$, the mixing angle for $h$ and $\Pi^6$ decreases with $y_{r2}$ and hence does not lead to a tension with the current Higgs particle measurement.

When the $\Pi^8$ quartic coupling is positive and stabilizes the vacuum,
 the physical mode of the $\Pi^8$ field, $\Pi^8_{{\rm phys}}=\Pi^8 - \langle \Pi^8 \rangle$, gains the following mass term:
\begin{align}
-{\cal L} &\supset \frac{1}{2} \, (-2 M^2_{88}) \, (\Pi^8_{{\rm phys}})^2 \equiv \frac{1}{2} \, M_{\Pi^8_{{\rm phys}}}^2 \, (\Pi^8_{{\rm phys}})^2.
\end{align}
Fortunately, the mass of the physical mode $M_{\Pi^8_{{\rm phys}}}$ is determined solely by $M^2_{88}$, with no dependence on the value of the quartic coupling.
We further stress that the mass of $\Pi^8_{{\rm phys}}$ is rather insensitive to $y_{r2}$;
 when $y_{r2}$ varies from $10^{-1}$ to $10^{-17}$, with $y_{r2}=10^{-17}$ corresponding to the case when $\Lambda_T$ is above the Planck scale
 $\Lambda_T \gtrsim 2.44\times10^{18}$~GeV,
 the mass changes in the following range:
\begin{align}
62~{\rm GeV} &< M_{\Pi^8_{{\rm phys}}} < 220~{\rm GeV} \ \ \ \ \ \ {\rm for \ } N=3 {\rm \ and \ } 10^{-1} < y_{r2} < 10^{-17},
\nonumber \\
41~{\rm GeV} &< M_{\Pi^8_{{\rm phys}}} < 155~{\rm GeV} \ \ \ \ \ \ {\rm for \ } N=6 {\rm \ and \ } 10^{-1} < y_{r2} < 10^{-17}.
\label{pi8physmass}
\end{align}
The insensitivity to $y_{r2}$ is because the mass term for $\Pi^8$ stems only from the $\bar{\chi}\psi H$ Yukawa interaction and is thus of the order of $y_{r2} \Lambda_T$.
However, the scale of $\Lambda_T$ is about the SM Higgs field mass divided by $y_{r2}$, \textit{i.e.}, $\Lambda_T \sim m_h/y_{r2}$, and therefore the $\Pi^8$ mass is of the same order as the SM Higgs field mass.
The mild logarithmic dependence on $y_{r2}$ is due to the fact that the SM Higgs particle mass functions as an infrared cutoff for the momentum integral in the ${\cal F}_{38}$ formula Eq.~(\ref{f3calc}).

It is convenient to organize pNG bosons in electroweak charge eigenstates as
 $\Pi = (\Pi^+,~\Pi^0,~\Pi^-) \equiv (\frac{1}{\sqrt{2}}(\Pi^1+i\Pi^2),~\Pi^3,~\frac{1}{\sqrt{2}}(\Pi^1-i\Pi^2))$,
 $\Sigma = (\Sigma^+,~\Sigma^0) \equiv (\frac{1}{\sqrt{2}}(\Pi^4+i\Pi^5),~\frac{1}{\sqrt{2}}(\Pi^6+i\Pi^7))$, $\bar{\Sigma} = (\bar{\Sigma}^0,~-\Sigma^-) = (\frac{1}{\sqrt{2}}(\Pi^6-i\Pi^7),~-\frac{1}{\sqrt{2}}(\Pi^4-i\Pi^5))$.
$\Pi$ is an isospin triplet with hypercharge $Y=0$, $\Sigma$ is an isospin doublet with hypercharge $Y=-1/2$, and $\Pi^8_{{\rm phys}}$ has no gauge charge.
Note that unlike neutral kaons, $\Sigma^0$ and $\bar{\Sigma}^0$ do not mix because of the absence of $\chi,\psi$ current mass.
The mass and electroweak charges of $\Pi$, $\Sigma$ and $\Pi^8_{{\rm phys}}$ are summarized in Table~\ref{png}.
\begin{table}[h]
\begin{center}
\begin{tabular}{|c|c|c|c|} \hline
               & $\Pi$ & $\Sigma$ & $\Pi^8_{{\rm phys}}$ \\ \hline
Mass           & $\sqrt{\dfrac{3}{N}}\dfrac{\sqrt{2}g_W}{y_{r2}}$ 25~GeV & $\sqrt{\dfrac{3}{N}}\dfrac{\sqrt{2g_W^2 + g_Z^2}}{2y_{r2}}$ 25~GeV & $ \sqrt{\dfrac{3}{N}} \sqrt{\log\left( 5.2 \, \dfrac{3}{N} \dfrac{1}{y^2_{r2}} \right)}$ 25~GeV \\ \hline   
$SU(2)_W \times U(1)_Y$ & \bf{3}$_{0}$ & \bf{2}$_{1/2}$ & \bf{1}$_{0}$ \\ \hline
\end{tabular}
\end{center}
\caption{
Mass and electroweak charges of the pNG bosons $\Pi$, $\Sigma$ and $\Pi^8_{{\rm phys}}$ in the limit with $y_{r2} \ll g_W$, where electroweak symmetry breaking is negligible.
That 25~GeV appears both in the $\Pi$, $\Sigma$ masses and in the $\Pi^8_{{\rm phys}}$ mass is merely accidental.
}
\label{png}
\end{table}
We find that $\Pi$ is always heavier than $\Sigma$, but the $\Pi$ mass does not exceed twice the $\Sigma$ mass.
$\Pi^8_{{\rm phys}}$ is the lightest pNG boson in any case.
\\

\subsection{Interactions, decay pattern and production channels of the pseudo-Nambu-Goldstone bosons}

The $\Pi$, $\Sigma$ and $\Pi^8$ fields couple to the physical Higgs field $h$, whose strength is determined by the following correlation functions:
\begin{align}
&\langle \, \Sigma \, \vert ( \, y \bar{\chi}_2\psi(0)+ y\bar{\psi}\chi_2(0) \, ) \vert \, \Pi \, \rangle, \ \ \ \ \ \langle \, \Pi^8 \, \vert ( \, y \bar{\chi}_2\psi(0) + y \bar{\psi}\chi_2(0) \, ) \vert \, \Sigma \, \rangle.
\end{align}
Evaluation of the above correlation functions is left for future works.
There is another type of interaction described by the Wess-Zumino-Witten (WZW) term~\cite{wz,w}
 in chiral perturbation theory for the pNG bosons $\Pi$, $\Sigma$ and $\Pi^8$.
This is particularly important for phenomenology of the $\Pi^8_{{\rm phys}}$ particle, as it is the principal source for the production and decay channels of the $\Pi^8_{{\rm phys}}$ particle.
We below extract phenomenologically relevant part of the WZW term~\cite{krs} which involves only $\Pi^8$ and electroweak gauge fields with the number of fields limited to 3 or 4:
\begin{align}
-{\cal L}_{WZW} &\supset \frac{N}{48\pi^2} \epsilon^{\mu\nu\rho\sigma} \frac{1}{f_\Pi} \, \left\{ \, 2\sqrt{3}g_W^2 \, \Pi^8 \partial_\mu W^+_\nu \partial_\rho W^-_\sigma + \sqrt{3}(1-2s_W^2+2b s_W^4)g_Z^2 \, \Pi^8 \partial_\mu Z_\nu \partial_\rho Z_\sigma \right.
\nonumber \\
&+ 2\sqrt{3}(1-2b s_W^2)g_Z \, e \, \Pi^8 \partial_\mu Z_\nu \partial_\rho A_\sigma + 2\sqrt{3}b \, e^2 \, \Pi^8 \partial_\mu A_\nu \partial_\rho A_\sigma 
\nonumber \\
&\left. -i\frac{1}{\sqrt{3}}g_W^3 \, W^+_\mu W^-_\nu (c_W Z_\rho + s_W A_\rho) \partial_\sigma \Pi^8 \, \right\},
\label{wzw}
\end{align}
 where $A_\mu$ denotes photon and $b/2$ is the hypercharge for the $\chi$ field.
We have the non-vanishing WZW term above because the axial current corresponding to $\Pi^8$ is anomalous with respect to the $SU(2)_W$ weak gauge group,
 and when $b\neq1/2$, also to the $U(1)_Y$ hypercharge gauge group.

The decay pattern of the pNG bosons is as follows.
For the $\Pi$ particle, since the $\Pi$ mass is below twice the $\Sigma$ mass, $\Pi \to \Sigma \bar{\Sigma}$ decay through the $SU(N)_T$ gauge interaction is kinematically forbidden.
The $SU(2)_W$ and $U(1)_Y$ current conservation prohibits $\Pi \to (W \, {\rm or} \, Z) + \Sigma$ and $\Pi \to (W \, {\rm or} \, Z) + \Pi^8_{{\rm phys}}$ decays.
Therefore, the main decay channels of the $\Pi$ particle are $\Pi^{\pm} \to h + \Sigma^{\pm}$, $\Pi^0 \to h + \Sigma^0$ and $\Pi^0 \to h + \bar{\Sigma}^0$ through the $\bar{\chi}\psi H$ Yukawa interaction,
 whose decay amplitudes are
\begin{align}
{\cal A}(\Pi^+(p+q) \to h(q) \, \Sigma^+(p)) &= \frac{1}{\sqrt{2}} \, \langle \, \Sigma^+(p) \, \vert y \, \bar{\chi}_2(0)\psi(0) \vert \, \Pi^+(p+q) \, \rangle,
\label{pidecay1} \\
{\cal A}(\Pi^0(p+q) \to h(q) \, \bar{\Sigma}^0(p)) &= \frac{1}{\sqrt{2}} \, \langle \, \bar{\Sigma}^0(p) \, \vert y \, \bar{\chi}_2(0)\psi(0) \vert \, \Pi^0(p+q) \, \rangle,
\label{pidecay2} \\
\left\vert {\cal A}(\Pi^-(p+q) \to h(q) \, \Sigma^-(p)) \right\vert &= \left\vert {\cal A}(\Pi^+(p+q) \to h(q) \, \Sigma^+(p)) \right\vert,
\nonumber \\
\left\vert {\cal A}(\Pi^0(p+q) \to h(q) \, \bar{\Sigma}^0(p)) \right\vert &= \left\vert {\cal A}(\Pi^0(p+q) \to h(q) \, \Sigma^0(p)) \right\vert,
\label{cp}
\end{align}
 where $p+q$ denotes the $\Pi$ meson momentum, $p_2$ denotes the final-state $\Sigma$ meson momentum, and $q$ satisfies $q^2=m_h^2$.
In addition to the above decay channel, $\Pi^{\pm} \to W^{\pm} \Pi^0$ or $\Pi^{0} \to W^{\mp} \Pi^{\pm}$ decay through the $SU(2)_W$ interaction is possible depending on the mass splitting in the $\Pi$ triplet, which we have omitted in the derivation of the mass spectrum.

For the $\Sigma$ particle,
 the $SU(2)_W$ and $U(1)_Y$ current conservation again forbids $\Sigma \to (W \, {\rm or} \, Z) + \Pi^8_{{\rm phys}}$ decays.
Hence, the $\Sigma^0$ and $\bar{\Sigma}^0$ particles dominantly decay as $\Sigma^0 \to h + \Pi^8_{{\rm phys}}$ and $\bar{\Sigma}^0 \to h + \Pi^8_{{\rm phys}}$ through the $\bar{\chi}\psi H$ Yukawa interaction,
 with the decay amplitudes below:
\begin{align}
{\cal A}(\Sigma^0(p+q) \to h(q) \, \Pi^8_{{\rm phys}}(p)) &= \frac{1}{\sqrt{2}} \, \langle \, \Pi^8(p) \, \vert y \, \bar{\psi}(0)\chi_2(0) \vert \, \Sigma^0(p+q) \, \rangle,
\label{sigmadecay} \\
\left\vert {\cal A}(\bar{\Sigma}^0(p+q) \to h(q) \, \Pi^8_{{\rm phys}}(p)) \right\vert &= \left\vert {\cal A}(\Sigma^0(p+q) \to h(q) \, \Pi^8_{{\rm phys}}(p)) \right\vert.
\label{cp2}
\end{align}
On the other hand, the only tree-level decay channel of the $\Sigma^{\pm}$ particle is $\Sigma^+ \to W^+ \bar{\Sigma}^0$ and $\Sigma^- \to W^- \Sigma^0$ through the $SU(2)_W$ interaction provided the mass splitting in the $\Sigma$ doublet allows it kinematically.
Otherwise, the main tree-level decay channel is $\Sigma^+ \to W^+ h \Pi^8_{{\rm phys}}$  $\Sigma^- \to W^- h \Pi^8_{{\rm phys}}$ via off-shell $\Sigma^0$ and $\bar{\Sigma}^0$ fields.

For the $\Pi^8_{{\rm phys}}$ particle, the decay proceeds through the WZW term Eq.~(\ref{wzw}).
The partial widths are calculated from Eq.~(\ref{wzw}) and are found to be (the latter three modes are valid only when kinematically allowed)
\begin{align}
\Gamma(\Pi^8_{{\rm phys}} \to \gamma \gamma) &= \frac{b^2 \, e^4}{3072\pi^5} \, N^2 \, \frac{M_{\Pi^8_{{\rm phys}}}^3}{f_\Pi^2} 
\nonumber \\
\Gamma(\Pi^8_{{\rm phys}} \to \gamma Z) &= \frac{e^2g_Z^2(1-2b \, s_W^2)^2}{6144\pi^5} \, N^2 \, \frac{M_{\Pi^8_{{\rm phys}}}^3}{f_\Pi^2} \left(1-\frac{M_Z^2}{M_{\Pi^8_{{\rm phys}}}^2}\right)^3 
\nonumber \\
&= \Gamma(\Pi^8_{{\rm phys}} \to \gamma \gamma) \, \frac{(1-2b \, s_W^2)^2}{2b^2 \, s_W^2 c_W^2}\left(1-\frac{M_Z^2}{M_{\Pi^8_{{\rm phys}}}^2}\right)^3,
\nonumber \\
\Gamma(\Pi^8_{{\rm phys}} \to Z Z) &= \frac{g_Z^4(1-2s_W^2 + 2b \, s_W^4)^2}{12288\pi^5} \, N^2 \, \frac{M_{\Pi^8_{{\rm phys}}}^3}{f_\Pi^2} \left(1-\frac{4M_Z^2}{M_{\Pi^8_{{\rm phys}}}^2}\right)^{3/2} 
\nonumber \\
&= \Gamma(\Pi^8_{{\rm phys}} \to \gamma \gamma) \, \frac{(1-2s_W^2 + 2b \, s_W^4)^2}{4b^2 \, s_W^4 c_W^4}\left(1-\frac{4M_Z^2}{M_{\Pi^8_{{\rm phys}}}^2}\right)^{3/2},
\nonumber \\
\Gamma(\Pi^8_{{\rm phys}} \to W^+ W^-) &= \frac{g_W^4}{6144\pi^5} \, N^2 \, \frac{M_{\Pi^8_{{\rm phys}}}^3}{f_\Pi^2} \left(1-\frac{4M_W^2}{M_{\Pi^8_{{\rm phys}}}^2}\right)^{3/2} 
\nonumber \\
&= \Gamma(\Pi^8_{{\rm phys}} \to \gamma \gamma) \, \frac{1}{2b^2 \, s_W^4}\left(1-\frac{4M_W^2}{M_{\Pi^8_{{\rm phys}}}^2}\right)^{3/2}.
\end{align}
Numerically, we have 
\begin{align}
\frac{1}{\Gamma(\Pi^8_{{\rm phys}} \to \gamma \gamma)} &= \frac{1}{b^2 N^2} \left(\frac{N}{N_c}\right)^{3/2} \frac{1}{y_{r2}^2} \frac{1}{\log^{3/2}\left( 5.2 \, \dfrac{3}{N} \dfrac{1}{y^2_{r2}} \right)} \, 1.48 \times 10^{-18} \ {\rm s},
\nonumber \\
\end{align}
Additionally, the $\Pi^8_{{\rm phys}} \to W^+ W^- \gamma$ decay is possible depending on the $\Pi^8_{{\rm phys}}$ mass, but the partial width is suppressed by the factor $e^2/(2\pi)^2$ compared to 
 the $\Pi^8_{{\rm phys}} \to W^+ W^-$ decay.

Note that the pNG bosons cannot decay into two SM fermions via a $s$-channel gauge boson,
 because $\chi$ and $\psi$ fields couple to the electroweak gauge fields through vector currents
\footnote{
In the current model, $\chi$ and $\psi$ must be vector-like with respect to the electroweak gauge groups in order to avoid electroweak symmetry breaking along
 chiral symmetry breaking.
},
 while the pNG bosons couple to axial-vector currents.

\begin{figure}[H]
  \begin{center}
    \includegraphics[width=80mm]{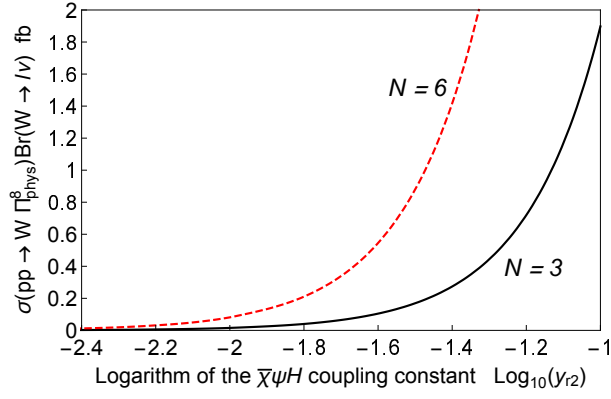}
    \caption{
    Cross section for the production of a $\Pi^8_{{\rm phys}}$ particle associated with a $W$ boson in 13~TeV $pp$ collisions, multiplied by the $W$ boson branching ratio into electron or muon.
    The horizontal axis is in the logarithm of $y_{r2}$, $\log_{10} y_{r2}$, and the solid and dashed lines respectively correspond to the cases with $N=3$ and $N=6$.
    }
    \label{w}
  \end{center}
\end{figure}

\begin{figure}[H]
  \begin{center}
    \includegraphics[width=80mm]{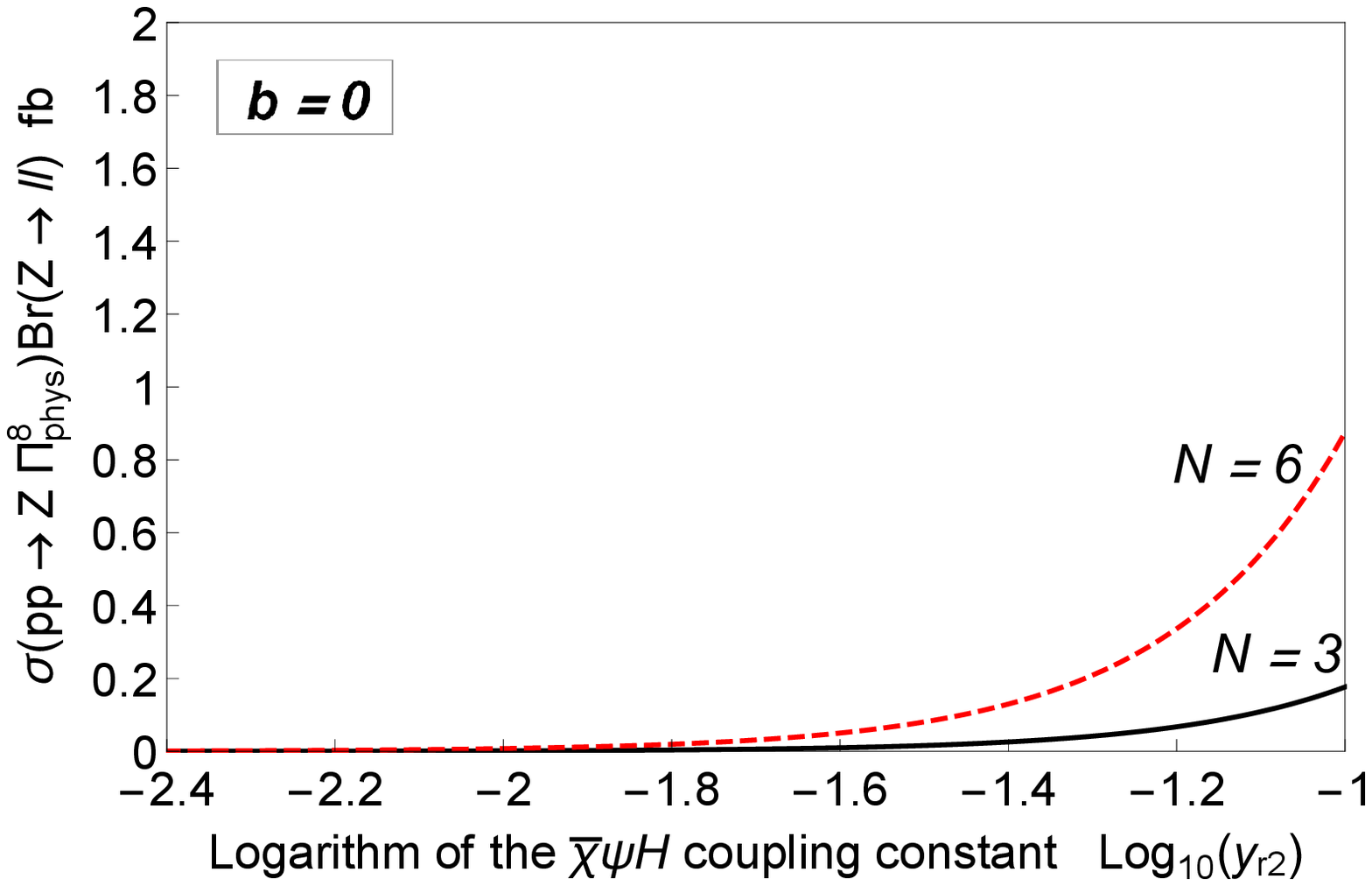}
    \includegraphics[width=80mm]{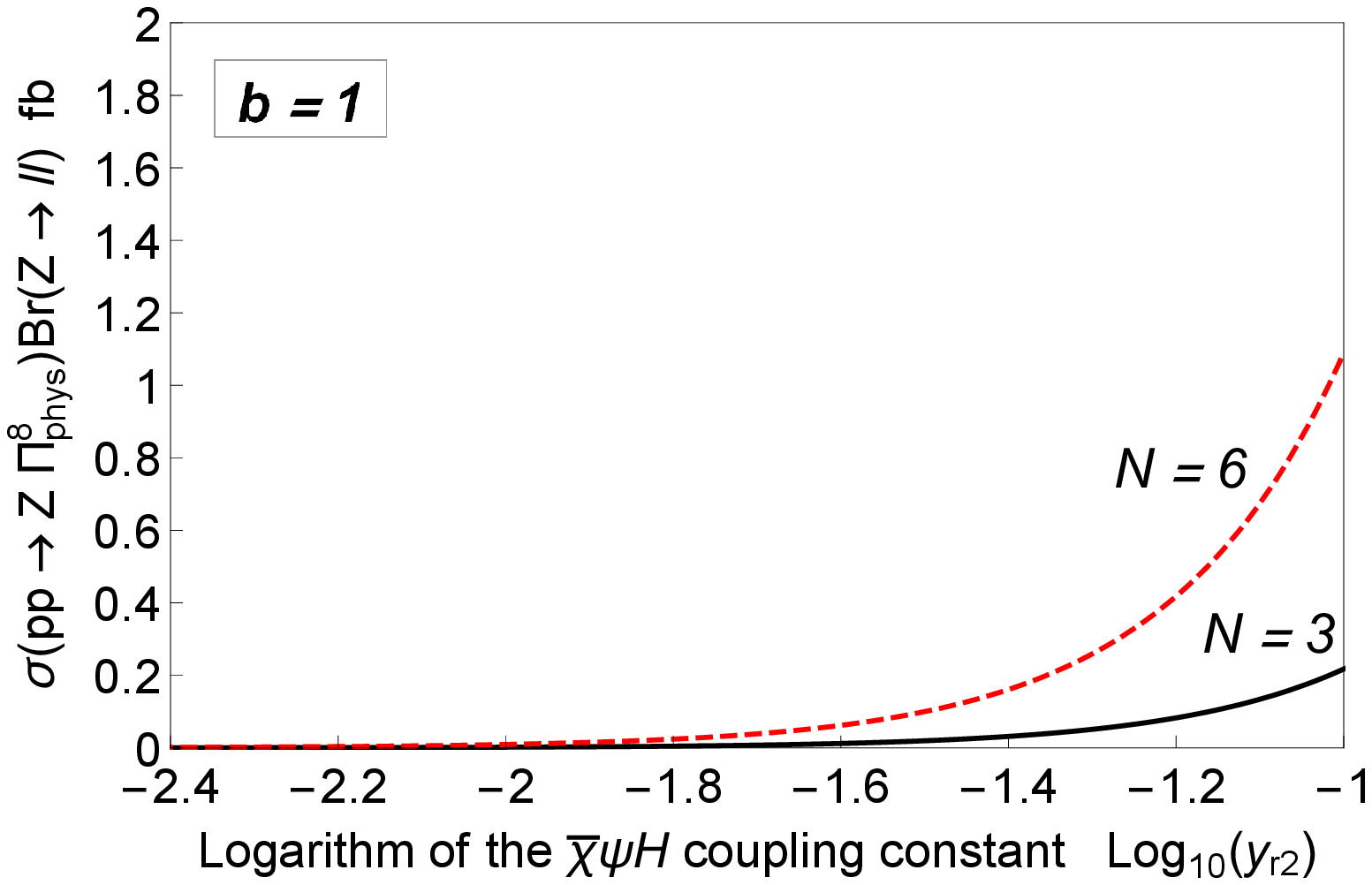}
    \caption{
    Cross section for the production of a $\Pi^8_{{\rm phys}}$ particle associated with a $Z$ boson in 13~TeV $pp$ collisions, multiplied by the $Z$ boson branching ratio into electrons or muons.
    The left plot is for $b=0$, with the hypercharge of $\chi$ being 0, and the right plot is for $b=1$, with the hypercharge of $\chi$ being $1/2$.
    The horizontal axis is in the logarithm of $y_{r2}$, $\log_{10} y_{r2}$, and the solid and dashed lines respectively correspond to the cases with $N=3$ and $N=6$.
    }
    \label{z}
  \end{center}
\end{figure}

\begin{figure}[H]
  \begin{center}
    \includegraphics[width=80mm]{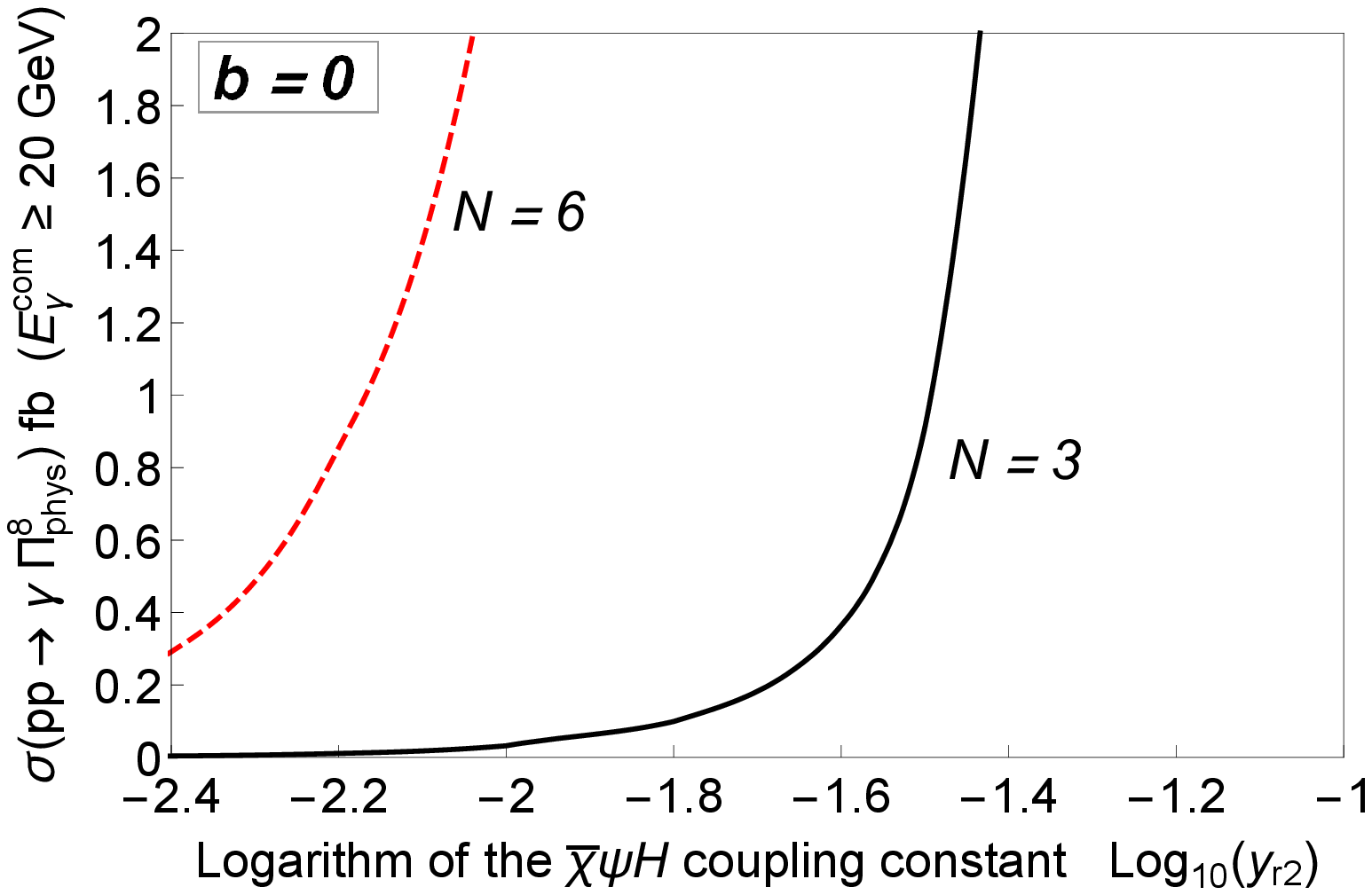}
    \includegraphics[width=80mm]{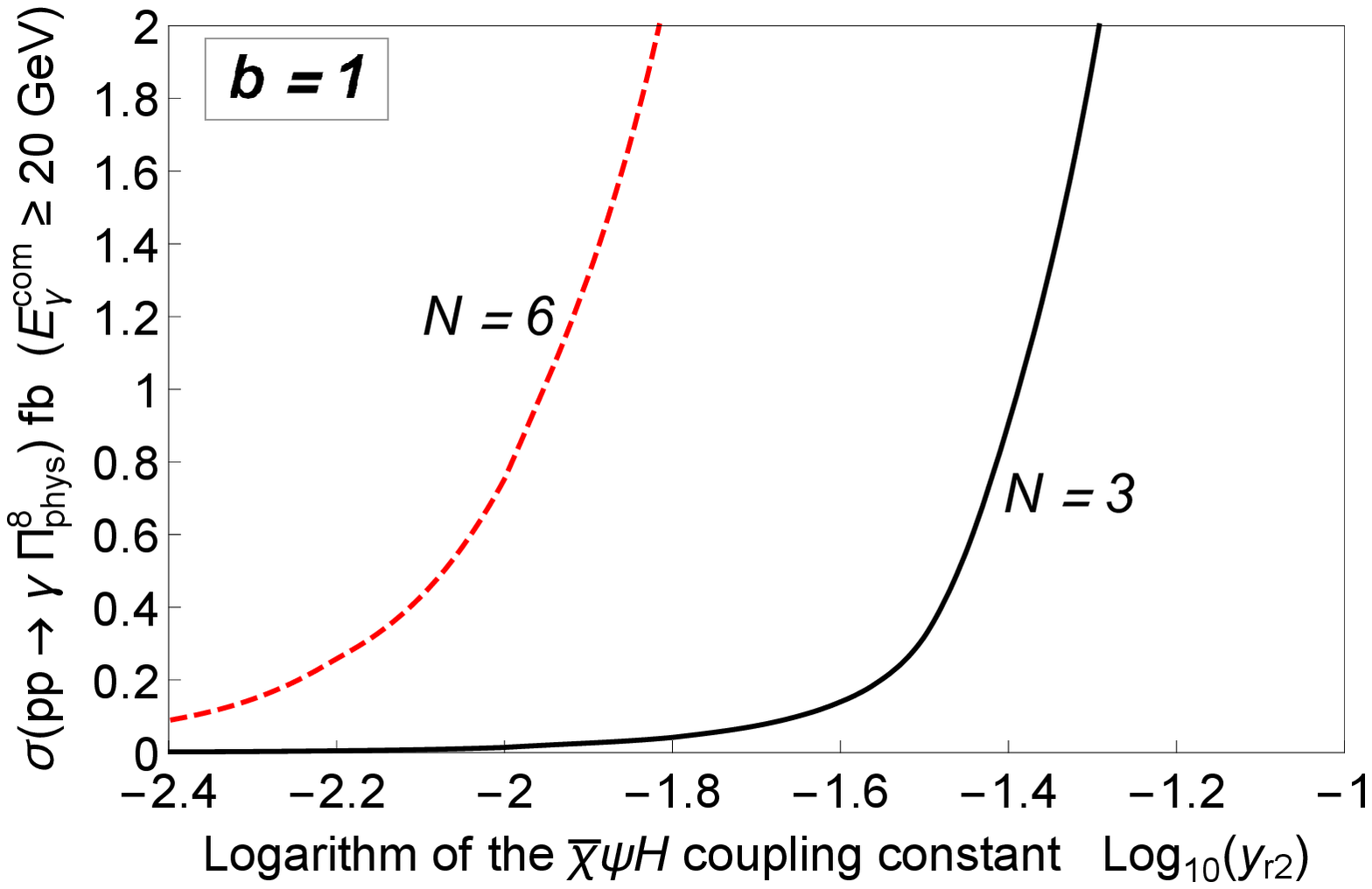}
    \caption{
    Cross section for the production of a $\Pi^8_{{\rm phys}}$ particle associated with a photon in 13~TeV $pp$ collisions,
     where the photon energy in the parton center-of-mass frame is above 20~GeV.
    The left plot is for $b=0$, with the hypercharge of $\chi$ being 0, and the right plot is for $b=1$, with the hypercharge of $\chi$ being $1/2$.
    The horizontal axis is in the logarithm of $y_{r2}$, $\log_{10} y_{r2}$, and the solid and dashed lines respectively correspond to the cases with $N=3$ and $N=6$.
    }
    \label{gam}
  \end{center}
\end{figure}

We briefly discuss the model's signatures at the LHC.
Since the $\Pi^8_{{\rm phys}}$ particle is predicted to have mass below 220~GeV, kinematically it is accessible at the LHC.
The main production channels are gauge-boson-associated productions through the WZW term Eq.~(\ref{wzw}) described as
\begin{align}
q\bar{q}' &\to W^* \to \Pi^8_{{\rm phys}} + W, \ \ \ \ \ q\bar{q} \to Z^*/\gamma^* \to \Pi^8_{{\rm phys}} + Z, \ \ \ \ q\bar{q} \to Z^{(*)}/\gamma^* \to \Pi^8_{{\rm phys}} + \gamma,
\nonumber \\
(&q, \, q' = u, \, d),
\end{align}
 where the intermediate gauge fields are off-shell except for the $\Pi^8_{{\rm phys}} + \gamma$ production, in which $Z$ boson can be on-shell.
Note that the vector boson fusion process does not enjoy collinear enhancement proportional to $\log(\hat{s}/M_{\Pi^8_{{\rm phys}}}^2)$
 ($\hat{s}$ denotes the parton center-of-mass energy), because the $\Pi^8$ field couples to the transverse polarization modes, not the longitudinal mode, of $W$ and $Z$ bosons.
Therefore, the vector-boson-fusion cross sections are simply suppressed by the electroweak gauge couplings and a phase space factor compared to those for the associated productions.
At hadron colliders, decay channels practically usable for detecting $\Pi^8_{{\rm phys}}$ particle signals are those in which 
 the $W$ or $Z$ boson associated with $\Pi^8_{{\rm phys}}$ decays into electron or muon, or the associated photon is not soft, namely,
\begin{align}
q\bar{q}' &\to W^* \to \Pi^8_{{\rm phys}} + W(\to \ell \nu), 
\ \ \ 
q\bar{q} \to Z^*/\gamma^* \to \Pi^8_{{\rm phys}} + Z(\to \ell \bar{\ell}),
\ \ \
q\bar{q} \to Z^{(*)}/\gamma^* \to \Pi^8_{{\rm phys}} + \gamma_{{\rm hard}}
\nonumber \\
(&\ell = e, \, \mu; \ \ \ \gamma_{{\rm hard}} {\rm \ denotes \ a \ photon \ sufficiently \ hard \ to \ be \ detected}).
\label{lhc}
\end{align}
This is because the presence of a muon, electron or photon associated with $\Pi^8_{{\rm phys}}$ decay products is indispensable for reducing SM backgrounds.
To make a crude estimate for the upper bound on $y_{r2}$ at the 13~TeV LHC, we present in Figures~\ref{w},~\ref{z},~\ref{gam} the cross section times branching ratio for
 each production and decay process in Eq.~(\ref{lhc}) in 13~TeV $pp$ collisions.
In the calculation, the leading order MSTW 2008 parton distribution function~\cite{mstw} is employed,
 with the factorization scale set at $\mu_F=M_{\Pi^8_{{\rm phys}}}+M_W$ for the $\Pi^8_{{\rm phys}} + W$ channel, 
 $\mu_F=M_{\Pi^8_{{\rm phys}}}+M_Z$ for the $\Pi^8_{{\rm phys}} + Z$ channel, and $\mu_F=M_{\Pi^8_{{\rm phys}}}$ for the $\Pi^8_{{\rm phys}} + \gamma$ channel.
Remind that since the $\Pi^8_{{\rm phys}}$ particle mass and the WZW term are entirely determined by the coupling constant $y_{r2}$ and the gauge group size $N$,
 the cross sections depend only on these variables.
In Figure~\ref{gam}, the energy of the associated photon in the parton center-of-mass frame is required to be above 20~GeV.
\noindent
In the range $10^{-2.4} < y_{r2} < 10^{-1}$ displayed above, the $\Pi^8_{{\rm phys}}$ particle mass varies as 88~GeV$> M_{\Pi^8_{{\rm phys}}} >$62~GeV for $N=3$ case,
 and 60~GeV$> M_{\Pi^8_{{\rm phys}}} >$41~GeV for $N=6$ case.
It follows that for $b\neq 0$, $\Pi^8_{{\rm phys}}$ almost exclusively decays into two photons
 and hence can be reconstructed as a diphoton resonance.
For $b=0$, the term $\Pi^8 \partial^\mu A^\nu \partial^\rho A^\sigma$ in the WZW term is zero and the decay into diphoton is absent,
 in which case $\Pi^8_{{\rm phys}}$ mainly decays into a photon and a pair of SM fermions via an off-shell $Z$ boson.
When the off-shell $Z$ boson decays into quarks, neutrinos or tau leptons, the reconstruction of a $\Pi^8_{{\rm phys}}$ particle is challenging. 
Although derivation of the precise bound on $y_{r2}$ calls for a collider simulation with rapidity and transverse momentum selection cuts,
 efficiency factor for the $\Pi^8_{{\rm phys}}$ particle reconstruction and examination of SM backgrounds,
 it is evident that the region with $y_{r2}\lesssim10^{-1.6}$ for $N=3$ case
 and that with $y_{r2}\lesssim10^{-2.4}$ for $N=6$ case would not be excluded even with about 10~fb$^{-1}$ of data at the 13~TeV LHC.

We finally comment on accessibility of the other pNG bosons, $\Pi$ and $\Sigma$, at the LHC.
Since the $\Pi$ and $\Sigma$ particles are charged under the electroweak gauge groups, they can be pair-produced through the Drell-Yan process.
When the upper bound on $y_{r2}$ is saturated with $N=3$ and $y_{r2} = 10^{-1.6}$, 
 the $\Pi$ and $\Sigma$ masses are found to be $M_\Pi = 920$~GeV and $M_\Sigma=590$~GeV,
 in which case $\Pi$ and $\Sigma$ particles may be detected at the 14~TeV LHC with 300~fb$^{-1}$ of data.
For $N=6$ and $y_{r2} = 10^{-2.4}$,
 the masses are found to be $M_\Pi = 4.1$~TeV and $M_\Sigma=2.6$~TeV, and $\Pi$ and $\Sigma$ particles are out of reach of the 14~TeV LHC.
\\

\section{Vanishing of the scalar quartic coupling at the Planck scale}

We demonstrate that the model can realize, with the top quark pole mass as large as 172~GeV, the vanishing of the scalar quartic coupling at the Planck scale $M_P=2.44\times10^{18}$~GeV,
 namely, a flat scalar potential at the Planck scale.
For this purpose, we evaluate the scalar quartic coupling at the Planck scale 
 by numerically solving the RG equations for the quartic coupling of the elementary scalar field $H$, the SM gauge couplings and top quark Yukawa coupling.
We vary the coupling constant $y_{r2}$ and the top quark pole mass, considering that the top quark pole mass is subject to sizable experimental and theoretical uncertainties.
We concentrate on the case with $N=3$ and the range $10^{-1.6} \geq y_{r2} \geq 10^{-8}$, where the upper limit is based on a rough experimental bound at the 13~TeV LHC estimated in Section~3.2.
Since $y_{r2}$ is that small, the $\bar{\chi}\psi H$ Yukawa interaction plays no role in the RG evolution of the scalar quartic coupling.
Still, it is an important parameter that determines at which scale the confinement occurs and the particle content changes from (SM particles+light pNG bosons) to (SM particles+fermionic particles made of $\chi,\psi$ fields).

To obtain the RG equations for our model, we alter the SM two-loop RG equations in Ref.~\cite{rge} by adding contributions of new particles.
We make the approximation that the particle content changes from (SM particles+light pNG bosons) to (SM particles+fermionic particles made of $\chi,\psi$ fields) 
 precisely at the energy scale $M_\Theta$ and ignore loop-level threshold corrections.
With this simplification, the new particle contributions are included in the following way:
Below the scale $\mu=M_\Theta$, the pNG bosons with electroweak charges, $\Pi$ and $\Sigma$, contribute to the RG runnings of the weak and hypercharge gauge couplings.
At the scale $\mu=M_\Theta$, the SM Higgs quartic coupling $\lambda^{SM}$ and top quark Yukawa coupling $y_t^{SM}$ are
 matched to the $H$ quartic coupling $\lambda$ and the Yukawa coupling $(Y_u)_{33}$ in Eq.~(\ref{lagrangian2}) by the conditions Eqs.~(\ref{matching2}),~(\ref{matching3}).
Above the scale $\mu=M_\Theta$, fermionic particles comprising $\chi,\psi$ fields contribute to the RG runnings of the weak and hypercharge gauge couplings.
Remember that $M_\Theta$ is linked to the coupling constant $y_{r2}$ as $M_\Theta = r \, m_{K_0^*(1430)} \simeq (191/y_{r2}) \, m_{K_0^*(1430)}$ through Eqs.~(\ref{mtheta}),~(\ref{ratio}).
We fix SM parameters as $M_W=80.384$~GeV, $\alpha_s(M_Z)=0.1184$ and $m_h=125.09$~GeV.
Also, the hypercharge of the $\chi$ field is set as $b/2=1/2$. However, we have separately confirmed that $O(1)$ variation of $b$ does not affect the main result.

\begin{figure}[H]
  \begin{center}
    \includegraphics[width=100mm]{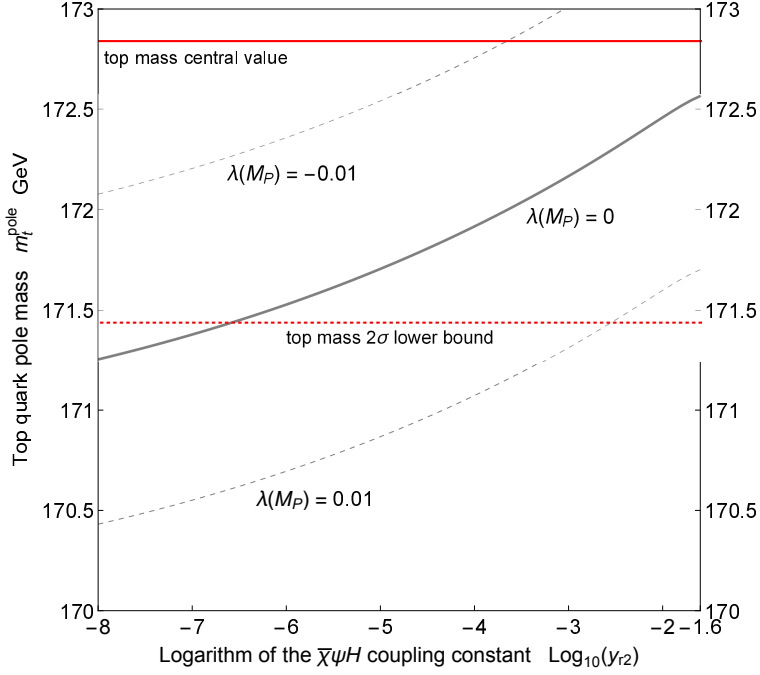}
    \caption{
    Contour plot of the quartic coupling for the elementary scalar field $H$ at the Planck scale, $\lambda(M_P)$.
    The $\bar{\chi}\psi H$ Yukawa coupling constant varies in the range $10^{-1.6} \geq y_{r2} \geq 10^{-8}$,
     and the top quark pole mass in the range $173~{\rm GeV} \geq m_t^{{\rm pole}} \geq 170~{\rm GeV}$.
    The thick curve corresponds to $\lambda(M_P)=0$, while the upper and lower dashed curves respectively correspond to $\lambda(M_P)=-0.01$ and $\lambda(M_P)=0.01$.
    Also shown are the central value (horizontal solid line) and the 2$\sigma$ lower bound (horizontal dashed line) for the top quark mass reported by the ATLAS Collaboration~\cite{topatlas}.
    }
    \label{lambda@planck}
  \end{center}
\end{figure}

\begin{figure}[H]
  \begin{center}
    \includegraphics[width=100mm]{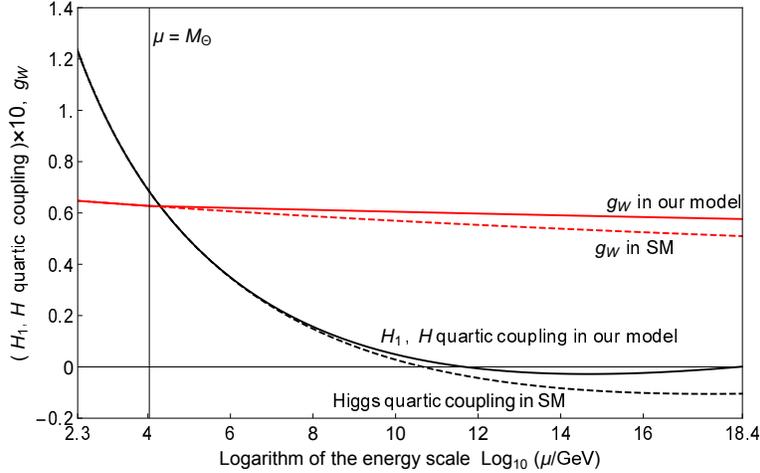}
    \caption{
    The RG evolutions of the $H_1$ and $H$ quartic coupling and weak gauge coupling $g_W$ in our model with $y_{r2}=10^{-1.6}$ and $m_t^{{\rm pole}}=172.55$~GeV,
     contrasted with those of the Higgs quartic coupling and weak gauge coupling in the SM with the same top quark pole mass.
    The scale $\mu$ moves from $\mu=10^{2.3}$~GeV$\simeq m_t^{{\rm pole}}$ to $\mu=10^{18.4}$~GeV$\simeq M_P$.
    The vertical line indicates the confinement scale $M_\Theta$.
    The quartic coupling of $H_1$ is shown below the scale $M_\Theta$, and that of $H$ is shown above that scale. 
    The solid black line starting from about 1.2 corresponds to the $H_1$ and $H$ quartic coupling, rescaled by $\times 10$ for visibility,
     and the solid red line corresponds to the weak gauge coupling.
    The nearby dashed lines describe the RG evolutions of the Higgs quartic coupling and weak gauge coupling in the SM.
    }
    \label{rge}
  \end{center}
\end{figure}

We present in Figure~\ref{lambda@planck} a contour plot of the value of the $H$ quartic coupling at the Planck scale, $\lambda(M_P)$,
 on the plane spanned by the coupling constant $y_{r2}$ and top quark pole mass $m_t^{{\rm pole}}$.
Additionally shown are the central value and the 2$\sigma$ lower bound for the top quark mass reported by the ATLAS Collaboration~\cite{topatlas}, which gives the top quark mass to be $m_t=172.84\pm0.70$~GeV.
\noindent
We find that $\lambda(M_P)=0$ can be achieved for the top quark pole mass as large as 172.5~GeV, which is quite consistent with the top quark mass reported by the ATLAS Collaboration.
This is a remarkable progress from previous models of classical scale invariance with $\lambda(M_P)=0$, where the top quark pole mass needs to be well below 172~GeV and hence outside the 2$\sigma$ bound.
The realization of $\lambda(M_P)=0$ with $m_t^{{\rm pole}}\simeq172.5$~GeV owes to the fact that fermionic particles made of $\chi$ field contribute positively to the RG running of the weak gauge coupling $g_W$ and thereby enhance it at high scales.
The weak gauge coupling then raises the $H$ quartic coupling through RG evolutions as compared to the SM, so that $\lambda(M_P)$ can reach 0 even when the top quark Yukawa coupling is large. 
The above situation is illustrated in Figure~\ref{rge},
 where we contrast the RG runnings of the scalar quartic coupling and the weak gauge coupling in our model with $y_{r2}=10^{-1.6}$ and $m_t^{{\rm pole}}=172.55$~GeV,
 with those of the Higgs quartic coupling and weak gauge coupling in the SM with the same top quark pole mass.
Here, the RG running of the $H_1$ quartic coupling is shown below the scale $M_\Theta$, and that of $H$ is shown above the scale $M_\Theta$.
\noindent
As $y_{r2}$ decreases, realization of $\lambda(M_P)=0$ requires smaller top quark pole mass.
This is because the pNG bosons, as being bosonic, contribute negatively to the RG evolution of the weak gauge coupling, in contrast to fermionic particles made of $\chi$ field.
Hence, the higher the confinement scale $M_\Theta$ is, the less the new particles enhance the weak gauge coupling and the scalar quartic coupling at ultraviolet scales.
Therefore, the top quark Yukawa coupling needs to be smaller to have $\lambda(M_P)=0$.
\\

\section{Summary and discussions}

We have studied a classically scale invariant extension of the Standard Model, in which 
 chiral symmetry breaking and confinement in a new $SU(N)_T$ gauge theory break the scale invariance.
The SM Higgs field emerges through the mixing of a scalar meson resulting from the confinement and an elementary scalar field.
The SM Higgs field mass is dynamically generated at the scale given by $\Lambda_T$, the dynamical scale of the $SU(N)_T$ gauge theory, times
 $y$, the Yukawa coupling constant for $SU(N)_T$-charged fermions and the elementary scalar field,
 automatically with the correct negative sign.
Concerning phenomenological signatures of the model, we have investigated the mass and interactions of pseudo-Nambu-Goldstone bosons associated with the spontaneous breaking of the $SU(3)_A$ axial symmetry along chiral symmetry breaking in the $SU(N)_T$ gauge theory.
We have found that the model predicts the existence of a Standard Model gauge singlet pseudoscalar particle with mass below 220~GeV, 
 which couples to two electroweak gauge bosons through the Wess-Zumino-Witten term, with the strength thus proportional to $1/\Lambda_T$.
Regarding the theoretical aspects, we have shown that the model can realize the vanishing of the scalar quartic coupling at the Planck scale
 with the top quark pole mass as large as 172.5~GeV, which is consistent with the current top quark mass measurement.

If $y_{r2}$ were as large as $0.3$, the gap between the Standard Model Higgs quartic coupling and the quartic coupling of $H$ as appears in Eq.~(\ref{matching2})
 would be sizable and could lead to the simultaneous vanishing of the scalar quartic coupling and its beta function at the Planck scale, realizing so-called "multiple-point principle" conjectured in Ref.~\cite{mpp}.
In the current model, however, $y_{r2}\simeq 0.3$ would incur too large cross section for the $\Pi^8_{{\rm phys}}$ particle production associated with a hard photon, which contradicts with the null result for new physics searches at the LHC.

We comment in passing that the $SU(N)_T$ gauge theory gives rise to baryonic states composed of $\chi,\psi$ fields.
The lightest baryon is stable due to the conservation of the $U(1)$ vector charge for $\chi,\psi$ fields.
We expect the lightest baryon to be a Standard Model gauge singlet, or at least neutral with respect to the electromagnetic interaction, because the electroweak interactions lift the baryon mass.
It can therefore be a dark matter candidate.
\\

\section*{Acknowledgement}

The authors thank Nobuchika Okada (University of Alabama), Noriaki Kitazawa (Tokyo Metropolitan University) and Masayasu Harada (Nagoya University) for useful comments.
This work is partially supported by Scientific Grants by the Ministry of Education, Culture, Sports, Science and Technology of Japan (Nos. 24540272, 26247038, 15H01037, 16H00871, and 16H02189).
\\

\end{document}